\documentclass[10pt]{article}
\pdfoutput=1
\usepackage{amssymb,amsmath,mathrsfs,enumerate}
\usepackage{epsf}
\usepackage{graphicx,rotate,multicol,slashed}
\usepackage{wrapfig}
\usepackage[dvipsnames]{xcolor}
\definecolor{mygreen}{RGB}{06,84,39}
\usepackage[colorlinks=true,
            linkcolor=mygreen,
            urlcolor=mygreen,
            citecolor=mygreen]{hyperref}
\usepackage{soul}
\usepackage{cite}
\usepackage{verbatim}
\usepackage{booktabs}
\usepackage[symbol]{footmisc}
\renewcommand*{\thefootnote}{\fnsymbol{footnote}}

\usepackage[capitalise]{cleveref}

\crefname{section}{Sec.}{Secs.}
\crefname{table}{Tab.}{Tabs.}
\crefname{figure}{Fig.}{Figs.}
\crefname{equation}{Eq.}{Eqs.}
\crefname{appendix}{Appendix\ }{Appendix\ }

\long\def\rpl#1!!#2!!{\textcolor{red}{#1} \textcolor{blue}{#2}}

\newcommand{\ba}{\begin{eqnarray}}
\newcommand{\ea}{\end{eqnarray}}
\newcommand{\be}{\begin{equation}}
\newcommand{\ee}{\end{equation}}

\usepackage[dvipsnames]{xcolor}

\newcommand{\U}[1]{\mathrm{U}(1)_{\mathrm{#1}}}			% Use this for U(1) groups
\newcommand{\SU}[2]{\mathrm{SU}(#1)_{\mathrm{#2}}}		% Use this for SU(N) groups

\def\ml{\mathscr L}

\def \order(#1){{\cal O} \left(#1 \right)}

\evensidemargin=\oddsidemargin
\def\Eqn#1{Eq.\ (\ref{#1})}
\def\Eqs#1#2{Eqs.\ (\ref{#1}) and (\ref{#2})}
\allowdisplaybreaks
\begin{document}

\begin{flushright}
   LU TP 19-12
\end{flushright}

\begin{center}
	{\Large \bf A three Higgs doublet model with symmetry-suppressed flavour changing neutral currents} \\
	\vspace*{1cm} {\sf Dipankar Das$^{a,}
        $\footnote{\href{mailto:d.das@iiti.ac.in}{\texttt{d.das@iiti.ac.in}}},~
        P.M. Ferreira$^{b,c,}$\footnote{\href{mailto:pmmferreira@fc.ul.pt}{\texttt{pmmferreira@fc.ul.pt}}},~
        Ant\'onio~P.~Morais$^{d,e,}$\footnote{\href{mailto:aapmorais@ua.pt}{\texttt{aapmorais@ua.pt}}},~
        Ian Padilla-Gay$^{f,}$\footnote{\href{mailto:ian.padilla@nbi.ku.dk}{\texttt{ian.padilla@nbi.ku.dk}}},~
        Roman Pasechnik$^{g,}$\footnote{\href{mailto:roman.pasechnik@thep.lu.se}{\texttt{roman.pasechnik@thep.lu.se}}},~
        J.~Pedro Rodrigues$^{d,e,}$\footnote{\href{mailto:joaopedrorodrigues@ua.pt}{\texttt{joaopedrorodrigues@ua.pt}}}} \\
	\vspace{10pt} {\small \em
    %%%%%%%%%%%%%%%%%%%%%%%%%%%%%%%%%%%%%%%%%%%%%%%%%%%%%%%%%%%%%
        $^a$Department of Physics, Indian Institute of Technology, Khandwa Road, Simrol, Indore 453 552, India\\
	    $^b$Instituto~Superior~de~Engenharia~de~Lisboa~---~ISEL, 1959-007~Lisboa, Portugal \\
        $^c$Centro~de~F\'{\i}sica~Te\'orica~e~Computacional, Faculdade~de~Ci\^encias,\\
        Universidade~de~Lisboa, Campo Grande, 1749-016~Lisboa, Portugal\\
        $^d$Departamento de F\'{i}sica da Universidade de Aveiro,
        Campus de Santiago, 3810-183 Aveiro, Portugal\\
        $^e$Centre  for  Research  and  Development  in  Mathematics  and  Applications  (CIDMA),\\
        Campus de Santiago, 3810-183 Aveiro, Portugal \\
        $^f$Niels Bohr International Academy and DARK, Niels Bohr Institute,\\ University of
        Copenhagen, Blegdamsvej 17, 2100, Copenhagen, Denmark\\
        $^g$Department of Astronomy and Theoretical Physics, Lund
        University, S\"{o}lvegatan 14A, 223 62 Lund, Sweden
 }
	\normalsize
\end{center}

%%%%%%%%%%%%%%%%%%%%%%%%%%%%%%%%%%%%%%%%%%%%%%%%%%%%%%%%%
%%%%  Changing footnote label back to arabic numbers %%%%
%%%%%%%%%%%%%%%%%%%%%%%%%%%%%%%%%%%%%%%%%%%%%%%%%%%%%%%%
\renewcommand*{\thefootnote}{\arabic{footnote}}
%%%%%%%%%%%%%%%%%%%%%%%%%%%%%%%%%%%%%%%%%%%%%%%%%%%%%%%
%%%%%  Resetting footnote counter  %%%%%%%%%%%%%%%%%%%
%%%%%%%%%%%%%%%%%%%%%%%%%%%%%%%%%%%%%%%%%%%%%%%%%%%%%%
\setcounter{footnote}{0}

\begin{abstract}
    \noindent
    We construct a three-Higgs doublet model with a flavour non-universal
    ${\rm U}(1)\times \mathbb{Z}_2$ symmetry. That symmetry induces
    suppressed flavour-changing interactions mediated by neutral
    scalars. New scalars with masses below the TeV scale can still
    successfully negotiate the constraints arising from flavour data. Such
    a model can thus encourage direct searches for extra Higgs bosons in the
    future collider experiments, and includes a non-trivial flavour structure.
\end{abstract}

\section{Introduction}
The properties of the new resonance observed at the LHC in 2012~\cite{Aad:2012tfa,Chatrchyan:2012xdj}
seem tantalizingly close to those of the Higgs boson predicted by the
Standard Model (SM) (for instance, see~\cite{Aad:2015zhl,Khachatryan:2016vau}).
The particle spectrum predicted by the SM has now been fully confirmed, but
many important questions in particle physics are left unanswered: the smallness of neutrino
masses, the fermion mass hierarchy, the colossal asymmetry between the quantity of
matter and antimatter in the universe and the nature of dark matter are a few of such unresolved issues. They are usually taken as hints for the existence of new physics (NP) beyond the SM (BSM). Typical BSM scenarios
that aim to fix one or more such shortcomings of the SM often end up
extending the scalar sector of the SM. In these extensions, the 125
GeV scalar observed at the LHC is not the only scalar in the spectrum
but the first one in a series of others to follow. This is an
intriguing possibility which motivates us for a closer inspection of
the properties of the observed scalar and inspires us to carry on our
efforts to look for new resonances at collider experiments.

When it comes to extending the scalar sector of the SM, adding replicas
of the SM Higgs-doublet is one of the simplest ways to do it. Such
extensions do not alter the tree level value of the electroweak (EW)
$\rho$-parameter. Two Higgs-doublet models (2HDMs), which add only one
extra doublet to the SM Higgs sector, have received its fair share of
attention through the years.
%PF
They were proposed by T.D. Lee in 1973~\cite{Lee:1973iz} as a means to obtain
a spontaneous breaking of the CP symmetry, and boast a rich phenomenology.
Other than the possibility of spontaneous CP breaking, such models contain
a richer particle spectrum, with a charged scalar and a total of three neutral ones,
may feature dark matter candidates in certain scenarios, and generically
give rise to the tree-level scalar-mediated flavour changing neutral currents (FCNCs).
Indeed, one ominous outcome of adding extra scalar doublets is that the fermions
of a particular charge will now receive their masses from
multiple Yukawa matrices and consequently, diagonalization of their mass
matrices will no longer guarantee the simultaneous diagonalization of the
Yukawa matrices. Therefore, in general, there will exist FCNCs at tree-level mediated by neutral scalars.

Experimental data from the flavour sector -- for instance,
neutral meson mass differences for Kaons and B-mesons, or $\epsilon_K$ data --
strongly constrain such FCNCs, typically forcing the extra scalars to have masses
above 1 TeV~\cite{Branco:1999fs}. An alternative is to fine-tune the FCNC interactions
so that they are very small, though for some models cancellations between CP-even and CP-odd
contributions to the amplitudes off the observables mentioned allow for below-TeV
scalars with a minimal fine-tuning (see, for
instance,~\cite{Ferreira:2011xc,Nebot:2015wsa,Ferreira:2019aps}). Another possibility
is to assume {\em alignment} between different Yukawa
matrices~\cite{Pich:2009sp,Jung:2010ik,Jung:2010ab,Egana-Ugrinovic:2018znw,Egana-Ugrinovic:2019dqu,Egana-Ugrinovic:2021uew}, though that is an {\em ansatz}
which is not preserved under renormalization~\cite{Ferreira:2010xe}. There is
yet another possibility, however: the BGL (Branco-Grimus-Lavoura)
model~\cite{Lavoura:1994ty,Branco:1996bq} is a remarkable
version of the 2HDM wherein FCNC interactions are {\em naturally} small -- this results
from the imposition of a flavour-violating symmetry, {\em i.e.} a symmetry which treats
some generations of fermions differently from others. As a consequence of this symmetry,
FCNC couplings are suppressed by off-diagonal Cabibbo-Kobayashi-Maskawa (CKM) matrix
elements. The phenomenology of the model has been studied quite thoroughly (see,
for instance,~\cite{Botella:2014ska,Botella:2015hoa,Bhattacharyya:2014nja}) and it remains a valid and
exciting possibility for BSM physics.

As a next-to-minimal possibility along these lines, recent years have seen a growing
interest in the topic of three Higgs-doublet models (3HDMs). These models add two extra
scalar doublets
%on and above
on top of the SM one, thereby conforming to the aesthetic appeal
of having three scalar generations in harmony with the three generations of fermions.
An early proposal by Weinberg~\cite{Weinberg:1976hu} included two discrete symmetries
which yielded flavour conservation. A recent study~\cite{Ferreira:2017tvy}, following
earlier work in~\cite{Ivanov:2011ae,Ivanov:2015mwl,Aranda:2016qmp}, used a generalized
CP symmetry (called CP4) to constrain the vast parameter space of 3HDM, and was found to
have large regions of parameter space which conformed with experimental constraints in
the flavour and scalar sectors. That model, however, used the liberty in parameters
present in the Yukawa sector to fit all observables, but the FCNC couplings were not necessarily small. Furthermore, it was shown in \cite{Ivanov:2021pnr} that the model contains several regions where top decays to light charged Higgs bosons is in conflict with LHC data. In \cite{Davoudiasl:2019lcg,Davoudiasl:2021syn} is was also studied how baryogenesis can be realized via the decay of new TeV-scale Higgs bosons in a 3HDM. Last but not least, in \cite{Darvishi:2021txa}, a maximally-symmetric version of the 3HDM based on the $\mathrm{Sp}(6)$ symmetry was proposed.

In this article, we will attempt
to implement the BGL method in a 3HDM framework. We construct a 3HDM
with a ${\rm U}(1)\times \mathbb{Z}_2$ flavor symmetry, where we also achieve the
smallness of the FCNC couplings in a way similar to the BGL case. In the considered
version of the model, tree-level FCNC interactions occur in the down-quark sector while
those in the up-quark sector are forbidden by the flavor symmetry. In this work,
we employ publicly available tools such as the generic BSM spectrum generator
\texttt{SARAH/SPheno} interfaced with the widely-used Higgs (\texttt{HiggsBounds/HiggsSignals})
and flavor (\texttt{Flavio}) observables' analysers enabling us to thoroughly verify
the model parameter space against the most relevant theoretical and experiments bounds. We find that relatively light scalars can successfully pass through the stringent experimental constraints arising from flavor data and hence may occur within the reach of the current and future collider experiments. To motivate
future searches, we also outline possible signatures of nonstandard scalars present in our model. For such a purpose we use \texttt{MadGraph5\_aMC@NLO 2.6.2} and focus on scalar and pseudoscalar production via gluon fusion with subsequent decay into tau leptons.

Our article is organized as follows. In \cref{sec:bgl} we review  the framework of BGL models. Then, in \cref{sec:mod}, we build our model, a 3HDM endowed with a ${\rm U}(1)\times \mathbb{Z}_2$ symmetry with a non-trivial structure in the Yukawa sector.
In \cref{sec:cons} we review the constraints imposed upon the model,
both theoretical -- boundedness from below, unitarity, electroweak precision bounds -- and experimental -- LHC Higgs data and searches for heavier scalars, flavour physics
data, among others. In \cref{sec:res} we explain in detail the procedure we followed
to perform a thorough numerical scan of the model and present the results we found for the parameter
space that survives all constraints imposed upon the model. We conclude in \cref{sec:conc}
with an overview of this work and a discussion of its significance.

\section{The BGL model}
\label{sec:bgl}

The BGL model is a version of the 2HDM where the scalar interactions with
fermions violate flavour -- meaning, unlike the interactions of the photon
and $Z$ boson, the neutral scalars in the BGL model ``jump families" like
the $W$ boson does. In 2HDM the general recipe to avoid FCNC is to only allow fermions of the same charge to couple
to just one of the doublets~\cite{Glashow:1976nt}. This is usually enforced by imposing a $\mathbb{Z}_2$ 
or U(1)~\cite{Peccei:1977hh} symmetry on the model (see also \cite{Campos:2017dgc,Arcadi:2021yyr}). 
The reason for doing this in the first place is the fact that tree-level mediated FCNCs would make significant contributions to flavour sector
observables such as the mass differences of the $K^0$, $B_d$ or $B_s$ mesons,
or to the $\epsilon_K$ quantity, ruining an agreement found within the SM for
those quantities -- unless the masses of the new scalars are all of order TeV,
or the FCNC Yukawa couplings are tuned to be very small.

The BGL model is remarkable since it forces the FCNC couplings to be heavily
suppressed as the result of a symmetry. The model therefore provides a simple
and natural explanation as to why NP contributions to flavour observables
would not ruin the agreement found within the SM, without the need of
any fine-tuning. To understand how this is achieved, consider the Yukawa Lagrangian
for the quark sector in the 2HDM,
\be
-\mathcal{L}_Y\, =\, \sum_{a,b=1}^3\,\left\{ \bar{Q}^a_L\,\left[ \left(\Gamma_1\right)_{ab} \phi_1 +
\left(\Gamma_2\right)_{ab} \phi_2\right] \,n_R^b\;+\;
\bar{Q}^a_L\,\left[ \left(\Delta_1\right)_{ab} \tilde{\phi}_1 +
\left(\Delta_2\right)_{ab} \tilde{\phi}_2 \right]\,p_R^b \right\}\,+\,\text{h.c.},
\label{eq:yuk}
\ee
where $Q^a_L = (p^a_L\, n^a_L)^T$ and $\phi_i$ are the SU(2) weak isospin (left-handed)
quark and Higgs doublets, respectively, and $\tilde{\phi}_k=i\sigma_2 \phi_k^*$. The $p$ and $n$ fields are the positively and negatively charged quark
fields, respectively. Upon rotation to the mass basis they will yield
the physical up and down quarks. The $a$ and $b$ are fermion family indices. $\Gamma$
and $\Delta$ are $3\times 3$ Yukawa coupling matrices for the down and up sector, respectively.
Upon spontaneous symmetry breaking, the scalar doublets develop neutral vacuum expectation values (VEVs),
such that\footnote{We are assuming
these VEVs are real which is the case of the BGL model but the generalization to complex ones is trivial.}
$\langle \phi_1\rangle = v_1/\sqrt{2}$ and $\langle \phi_2\rangle = v_2/\sqrt{2}$, with
$v_1^2 + v_2^2 = (246\text{ GeV})^2$. We define $\tan\beta = v_2/v_1$. For a CP-conserving model
(both at the explicit and vacuum levels), the model will have a charged scalar $H^+$, a pseudoscalar $A$ and
two CP-even scalars, $h$ and $H$. The $2\times 2$ CP-even mass matrix is diagonalized via an angle $\alpha$.

The up and down quark mass matrices are then given by
\be
M_p \,=\,\frac{1}{\sqrt{2}}\,(\Delta_1 v_1 + \Delta_2 v_2)\;\;\; , \;\;\;
M_n \,=\,\frac{1}{\sqrt{2}}\,(\Gamma_1 v_1 + \Gamma_2 v_2)\,,
\ee
the eigenvalues of which will be the physical quark masses. In fact, these mass matrices will be
bidiagonalized in the usual form as
\be
    D_u = V_L^\dagger M_p V_R = {\rm diag}\{m_u,m_c,m_t\} \;\;\;, \;\;\;
    D_d = U_L^\dagger M_n U_R = {\rm diag}\{m_d,m_s,m_b\} \,,
\label{e:dudd}
\ee
where $m_x$ are the physical quark masses, whereas $V$ and $U$ are U(3) matrices. These matrices relate the physical quark states $u$ and $d$ to the $p$ and $n$ original states in the
 following manner:
\ba
p_L = V_L \, u_L \,, && p_R = V_R \, u_R \,, \nonumber \\
n_L = U_L d_L \, \,, && n_R = V_R \, d_R \,.
\label{eq:qrot}
\ea
The CKM matrix is then obtained as
\begin{eqnarray}
\label{e:CKM}
    V = V_L^\dagger U_L \,.
    \label{eq:CKM}
\end{eqnarray}
We also  define the following matrices,
\be
N_u \,=\,\frac{1}{\sqrt{2}}\,V^\dagger_L\,(\Delta_1 v_2 - \Delta_2 v_1)\,V_R\;\;\; , \;\;\;
N_d \,=\,\frac{1}{\sqrt{2}}\,U^\dagger_L\,(\Gamma_1 v_2 - \Gamma_2 v_1)\,U_R\,,
\ee
which end up being related to the Yukawa couplings between the physical scalars and quarks. In fact,
with the usual conventions (see for instance~\cite{Branco:2011iw,Ferreira:2019aps}), the Yukawa
Lagrangian for the physical fields may be written as
\ba
-\mathcal{L}_\mathrm{Y} &=&
\frac{iA}{v}\, \left[ \bar u \left( N_u P_R - N_u^\dagger P_L \right) u
+ \bar d \left( N_d^\dagger P_L - N_d P_R \right) d \right]
\nonumber \\ & &
+ \frac{h}{v}\, \bar u \left[
  \left( s_{\beta - \alpha} M_u - c_{\beta - \alpha} N_u^\dagger \right) P_L
  + \left( s_{\beta - \alpha} M_u - c_{\beta - \alpha} N_u \right) P_R
  \right] u
\nonumber
\\ & &
+ \frac{h}{v}\, \bar d \left[
  \left( s_{\beta - \alpha} M_d - c_{\beta - \alpha} N_d^\dagger \right) P_L
  + \left( s_{\beta - \alpha} M_d - c_{\beta - \alpha} N_d \right) P_R
  \right] d
\nonumber \\ & &
+ \frac{H}{v}\, \bar u \left[
  \left( c_{\beta - \alpha} M_u + s_{\beta - \alpha} N_u^\dagger \right) P_L
  + \left( c_{\beta - \alpha} M_u + s_{\beta - \alpha} N_u \right) P_R
  \right] u \nonumber
\\ & &
+ \frac{H}{v}\, \bar d \left[
  \left( c_{\beta - \alpha} M_d + s_{\beta - \alpha} N_d^\dagger \right) P_L
  + \left( c_{\beta - \alpha} M_d + s_{\beta - \alpha} N_d \right) P_R
  \right] d
\nonumber \\ & &
+ \frac{\sqrt{2} H^+}{v}\,
\bar u \left( N_u^\dagger V P_L - V N_d P_R \right) d
+ \frac{\sqrt{2} H^-}{v}\, \bar d \left( V^\dagger N_u P_R
- N_d^\dagger V^\dagger P_L \right) u\,,
\label{eq:2hdm}
\ea
where we used the notation $s_x \equiv \sin x$, $c_x \equiv \cos x$.
On a side note, we can see from the above Lagrangian how in the {\em alignment limit} the lighter
Higgs' Yukawa interactions are exactly like those of the SM particles: in that limit one
has $\sin(\beta - \alpha) = 1$ -- which forces the vertices between $h$ and the electroweak
gauge bosons to be identical to those of the SM Higgs particle -- and therefore the
Yukawa couplings of $h$ to quark pairs are proportional to the quark mass, since the
contribution of the $N$ matrices vanishes.

In models with flavour conservation, each family of fermions of the same electric charge couples to a single Higgs doublet, via the imposition of discrete $\mathbb{Z}_2$~\cite{Glashow:1976nt,Paschos:1976ay}
or global U(1)~\cite{Peccei:1977hh} symmetries. Then, the diagonalization of the $M_u$ and $M_d$ matrices, \cref{e:dudd}, is the same as that of matrices $N_u$ and $N_d$ and
there are no flavour-violating Yukawa interactions mediated by neutral scalars. In general,
though, that will not be the case and FCNCs occur at tree level.

The BGL model is based on a symmetry imposed on the whole of the Lagrangian, where some of the quark and scalar fields transform as
\be
Q_{L1} \to e^{i \theta} Q_{L1}, \quad
p_{R1} \to e^{2 i \theta} p_{R1}, \quad
\Phi_2 \to e^{i \theta} \Phi_2,
\label{eq:sym}
\ee
with $\theta\neq n\pi$, with $n$ an arbitrary integer. All other fields remain
invariant under this symmetry. As we see, the symmetry treats differently one of the
generations of quarks~\footnote{In fact, these are unrotated quark fields, not yet
corresponding to physical quarks, but the principle holds.}, since only the ``first
family'' of quarks is affected by the transformations above. In fact, there are six
(not counting the leptonic sector)
models of the BGL type, which depend on which generation of quarks is chosen in \cref{eq:sym} above. For the scalar sector, the above symmetry transformation yields a Peccei-Quinn~\cite{Peccei:1977hh} scalar potential, which must be complemented with a soft breaking parameter as to yield a massive pseudoscalar particle.

In the quark sector, the symmetry transformations of \cref{eq:sym} impose zero values in several entries of the Yukawa matrices of \cref{eq:yuk}. They are found to be
\ba
\Gamma_1 = \left( \begin{array}{ccc}
0 & 0 & 0 \\ \times & \times & \times \\ \times & \times & \times
\end{array} \right),
 & &
\Gamma_2 = \left( \begin{array}{ccc}
\times & \times & \times \\ 0 & 0 & 0 \\ 0 & 0 & 0
\end{array} \right)\,,\nonumber \\
\quad
\Delta_1 = \left( \begin{array}{ccc}
0 & 0 & 0 \\ 0 & \times & \times \\ 0 & \times & \times
\end{array} \right),
 & &
\Delta_2 =\left( \begin{array}{ccc}
\times & 0 & 0 \\ 0 & 0 & 0 \\ 0 & 0 & 0
\end{array} \right)\,,
\label{eq:bgld}
\ea
where, in general, ``$\times$" indicates a complex non-zero entry. Then, the form
of $\Delta_1$ and $\Delta_2$ implies that the matrix $M_p$ is block
diagonal, namely
\be
M_p = \left( \begin{array}{ccc}
\times & 0 & 0 \\
0 & \times & \times \\ 0 & \times & \times
\end{array} \right).
\ee
and it may be bi-diagonalized by unitary matrices $V_L$ and $V_R$ of the form
\be
V_L = \left( \begin{array}{ccc}
1 & 0 & 0 \\ 0 & \times & \times \\ 0 & \times & \times
\end{array} \right),
\quad
V_R = \left( \begin{array}{ccc}
e^{ i \theta_R} & 0 & 0 \\
0 & \times & \times \\ 0 & \times & \times
\end{array} \right)\,.
\label{eq:VL}
\ee
with some phase $\theta_R$. The shape of $V_L$ is crucial for the FCNC suppression. First, though, we see that the matrices $V_L$ and $V_R$ can simultaneously bi-diagonalize
$\Delta_1$ and $\Delta_2$, and therefore $M_u$ and $N_u$ are both diagonal in the basis of physical up quarks. A simple calculation yields
\be
N_u\,=\,\mathrm{diag}\left(-\frac{m_{u_1}}{\tan\beta}\,,\,m_{u_2}\,\tan\beta\,,\,m_{u_3}\,\tan\beta\right)\,,
\ee
with $m_{u_{1,2,3}}$ the masses of the up-type quarks. However, we have not yet specified which generation is affected by the BGL transformations. Nevertheless, the above shows that for any choice of quark generation in \cref{eq:sym} the $N_u$ matrix is diagonal
in the up-type quark mass basis, and therefore -- since this matrix contains the quark Yukawa couplings of the neutral scalars -- {\em no FCNC occurs in the up sector}.

The down-quark sector is a different story: given the form of the $V_L$ matrix in
Eq.~\eqref{eq:VL} and with the CKM matrix $V$ given by~\eqref{eq:CKM}, we
immediately obtain
\be
U_L \equiv \left( \begin{array}{ccc}
V(1,1) & V(1,2) & V(1,3) \\ \times & \times & \times \\ \times & \times & \times
\end{array} \right)\,,
\ee
and the impact of this structure on the left rotation matrix for the down-quark sector is
considerable. Indeed, a straightforward calculation yields, for the $N_d$ matrix (which, we
remind the reader, contains the Yukawa couplings between the physical down-type quarks
and scalars)
\be
\left( N_d \right)_{aa} = m_a \left( \tan{\beta} -
\frac{\left| V_{1a} \right|^2}{\sin{\beta} \cos{\beta}} \right) \,,
\ee
for the diagonal terms, whereas the off-diagonal ones are given by
\be
\left( N_d \right)_{ab} = - \frac{V_{1a}^\ast V_{1b}}{\sin{\beta} \cos{\beta}}\,
m_b \quad (a \neq b)\,.
\ee
Thus we see that the off-diagonal Yukawa couplings between scalars and down-type
quarks -- which determine the strength of the FCNC interactions -- are CKM-suppressed. There is a freedom to choose the ``first'' family as any one of the physical quark generations,
and therefore one has three BGL models with FCNC in the down-quark sector and without them in the up-sector. An analogous symmetry to that of \cref{eq:sym} associated with a given family
of up-type quarks would yield other three models, where CKM-suppressed FCNC occur in the up-quark sector and where the down-sector is free from such flavour violation interactions.

This then is how the hallmark of the BGL models is achieved: a flavour-breaking
symmetry, which yields off-diagonal FCNC couplings naturally suppressed by the entries of the CKM matrix
elements. In what follows we build a similar model but for the case of three Higgs doublets.

%%%%%%%%%%%%%%%%%%%%%%%%%%%%%%%%%%%%%%%%%%%%%%%%%%%%%%%%%%
\section{A BGL-like 3HDM}
\label{sec:mod}

Beyond the aesthetic reason of considering three Higgs doublets in analogy with three fermion families, or the intellectual challenge of attempting to reproduce the BGL structure with a larger scalar sector (see~\cite{Botella:2009pq} for an earlier attempt), there are other reasons to explore a 3HDM with similarly suppressed FCNCs. The BGL model is quite successful, but recent studies~\cite{Botella:2014ska} have found that its parameter space may be quite constrained. A possible criticism one may levy at the analysis of~\cite{Botella:2014ska} is that the latter has extended the BGL structure to the leptonic sector as well -- something that is not mandatory as the model has enough freedom to
accommodate a flavour-preserving leptonic sector in what concerns the Yukawa interactions -- which is what we will consider here. Nonetheless, this shows that even with natural FCNC coupling suppression via off-diagonal CKM matrix elements, the BGL structure can be quite constrained from experimental data. Working within the framework of
a 3HDM will in principle imply greater freedom in terms of parameters that can be adjusted to comply with experimental bounds.

There is also another reason, more theoretical and fundamental, to attempt a 3HDM study of the BGL paradigm. In many instances, comparisons of the 2HDM with 3HDMs have revealed how special a model the 2HDM is. To give only a few examples, tree-level vacuum stability against charge breaking or spontaneous CP breaking was found for charge-and-CP conserving minima within the 2HDM~\cite{Ferreira:2004yd,Barroso:2005sm,Barroso:2007rr}, but charge breaking minima were found to coexist with charge-preserving ones for NHDMs with $N\geq 3$~\cite{Barroso:2006pa}; a full listing of all possible symmetries of the ${\rm SU}(2)\times {\rm U}(1)$ invariant 2HDM was found~\cite{Ivanov:2006yq,Ivanov:2007de} while for 3HDM we refer to~\cite{Ivanov:2011ae,Ivanov:2014doa,Darvishi:2019dbh}; generic bounded-from-below~\cite{Ivanov:2006yq,Ivanov:2007de} and unitarity~\cite{Ginzburg:2005dt} bounds were found for the 2HDM, but for the 3HDM such bounds only exist for particular versions of the model. As such, the possibility of ascertaining whether the BGL structure can be extended to a full 3HDM compels us to try to find it. And of course one can obtain an {\em exact} BGL-like 3HDM -- all one needs
to do is copying the procedure detailed in the previous section for the first two doublets,
while keeping the third doublet VEVless. For that to happen one would
consider the construction of the Inert Doublet Model (IDM)~\cite{Ma:1978,Barbieri:2006,Cao:2007,LopezHonorez:2006} and impose a $\mathbb{Z}_2$ symmetry on the third doublet, $\phi_3 \rightarrow -\phi_3$. This model would have FCNCs in the visible sector and also a dark matter candidate stemming from the third doublet. An interesting model would be recovered, however it would not solve our aim to
obtain a larger freedom to fit the flavour observables in comparison to what one has in the 2HDM BGL model. Thus we are compelled to consider a 3HDM wherein all doublets acquire VEVs.

Let us start by describing the scalar sector of the model, which
contains three spin-0 $\SU{2}{}$ doublets, $\phi_1$, $\phi_2$ and $\phi_3$.

\subsection{The scalar sector}
\label{sec:sca}
The scalar doublets are made to transform under the ${\rm U}(1)\times \mathbb{Z}_2$
symmetry as follows:
\begin{subequations}
	\label{e:stran}
	\begin{eqnarray}
	{\rm U}(1) &:& \phi_1 \to e^{i\alpha} \phi_1 \,, \qquad
	\phi_3 \to e^{i\alpha} \phi_3 \,. \\
	\mathbb{Z}_2 &:& \phi_1 \to - \phi_1 \,, \qquad
	\phi_2 \to \phi_2 \,, \qquad \phi_3 \to \phi_3 \,.
	\end{eqnarray}
\end{subequations}
We further require the scalar potential to be CP-invariant, {\em i.e} to be invariant under the usual CP transformations,
\be
\phi_1 \to \phi_1^*\,, \qquad \phi_2 \to \phi_2^*\,, \qquad \phi_3 \to \phi_3^*\,,
\ee
such that it can be written as
\begin{eqnarray}
V_0(\phi_1,\phi_2,\phi_3) &=&
\mu_1^2 \left(\phi_1^\dagger\phi_1 \right) + \mu_2^2 \left(\phi_2^\dagger\phi_2 \right) + \mu_3^2 \left(\phi_3^\dagger\phi_3 \right)
+\lambda_1 \left(\phi_1^\dagger\phi_1 \right)^2
\nonumber \\
&& +\lambda_2 \left(\phi_2^\dagger\phi_2 \right)^2 +\lambda_3 \left(\phi_3^\dagger\phi_3 \right)^2 +\lambda_4 \left(\phi_1^\dagger\phi_1 \right) \left(\phi_2^\dagger\phi_2 \right) +\lambda_5 \left(\phi_1^\dagger\phi_1 \right) \left(\phi_3^\dagger\phi_3\right)
\nonumber \\
&& +\lambda_6 \left(\phi_2^\dagger\phi_2\right) \left(\phi_3^\dagger\phi_3\right) +\lambda_7 \left(\phi_1^\dagger\phi_2\right) \left(\phi_2^\dagger\phi_1\right)
+\lambda_8 \left(\phi_1^\dagger\phi_3\right) \left(\phi_3^\dagger\phi_1\right) +\lambda_9 \left(\phi_2^\dagger\phi_3\right) \left(\phi_3^\dagger\phi_2\right)
\nonumber \\
&& +\lambda_{10} \left\{\left(\phi_1^\dagger\phi_3 \right)^2 + {\rm h.c.} \right\} \,.
\label{e:pot}
\end{eqnarray}
The most general potential that softly breaks the ${\rm U}(1)\times \mathbb{Z}_2$ flavor symmetry is given by
\begin{eqnarray}
V_{\rm soft}(\phi_1,\phi_2,\phi_3) = \mu_{12}^2 \phi_1^\dagger\phi_2 + \mu_{13}^2 \phi_1^\dagger\phi_3 +
\mu_{23}^2 \phi_2^\dagger\phi_3 + {\rm h.c.} \,, \qquad V = V_0 + V_{\rm soft} \,.
\label{e:pot-soft}
\end{eqnarray}
Here, due to the CP symmetry all the parameters in $V=V(\phi_1,\phi_2,\phi_3)$ are real. We have introduced real soft
breaking terms, in $V_{\rm soft}$, to avoid the appearance of a massless axion in the physical spectrum. Recall that the same procedure was necessary for the 2HDM BGL\cite{Bhattacharyya:2013rya,Bhattacharyya:2015nca}, due to the analogous breaking of the U(1) symmetry given by \cref{eq:sym}.

After spontaneous symmetry breaking, all doublets acquire real VEVs~\footnote{It is obviously possible to obtain spontaneous CP violation within a 3HDM, via complex VEVs, but we do not consider that more generic case in the current work.} and are expanded as
\be
\phi_k = \begin{pmatrix}
	w_k^+ \\ \frac{1}{\sqrt{2}} (v_k+h_k+iz_k)
\end{pmatrix} \,, \qquad (k=1,2,3) \,,
\label{eq:doub}
\ee
where $v_k$ represent the VEVs of each doublet which satisfy $v_1^2 + v_2^2 + v^2_3= (246\, \text{GeV})^2$.
The minimization of the potential yields three equations that can be conveniently resolved by expressing ~the quadratic mass parameters $\mu_1^2$, $\mu_2^2$ and $\mu_3^2$ in terms of the
three VEVs and other couplings as follows:

\begin{subequations}
	\begin{eqnarray}
	\mu_1^2 & =& -\frac{1}{2} \left[2 \lambda _1 v_1^2+\left(\lambda _4+\lambda _7\right) v_2^2+\left(\lambda _5+\lambda _8+2 \lambda _{10}\right) v_3^2+\frac{2 \left(\mu _{13}^2 v_3+\mu_{21}^2
		v_2\right)}{v_1}\right] \,, \\
	\mu_2^2 & =& -\frac{1}{2} \left[ 2 \lambda _2 v_2^2 + \left(\lambda _4+\lambda _7\right) v_1^2+\left(\lambda _6+\lambda _9\right) v_3^2+\frac{2 \left(\mu _{21}^2 v_1+\mu _{23}^2 v_3\right)}{v_2}\right] \,, \\
	\mu_3^2 & =& -\frac{1}{2}\left[2 \lambda _3 v_3^2+\left(\lambda _6+\lambda _9\right) v_2^2+\left(\lambda _5+\lambda _8+2 \lambda _{10}\right) v_1^2+\frac{2 \left(\mu _{13}^2 v_1+ \mu _{23}^2 v_2 \right)}{v_3} \right] \,.
	\end{eqnarray}
\end{subequations}

For latter use, we parameterize the VEVs as,
\begin{equation}
\label{Eq:VeVs_Sines/Cosines}
v_1 = v\sin\beta_1 \cos\beta_2 \,, \qquad v_2= v \sin\beta_2 \,, \qquad
v_3 = v\cos\beta_1 \cos\beta_2  \,, \qquad
v=\sqrt{v_1^2+v_2^2+v_3^2}
\end{equation}
and setting $v_{13}=\sqrt{v_1^2 + v_3^2}$, define the following orthogonal matrix which rotates the gauge eigenstates into the so-called Higgs basis, greatly simplifying the analysis of the scalar sector,
\begin{equation}
%\label{Eq:Ians(real)VEVMatrix}
\mathcal{O}_\beta = \left(
\begin{array}{ccc}
v_1/v & v_2/v & v_3/v \\
    v_3/v_{13} & 0  &  -v_1/v_{13}  \\
    v_1 v_2/(v v_{13}) & -v_{13}/v & v_2 v_3/(v v_{13}) \\
\end{array}
\right)
=
\left(
\begin{array}{ccc}
\sin\beta_1 \cos\beta_2 & \sin\beta_2 & \cos\beta_1 \cos\beta_2  \\
    \cos\beta_1 & 0  &  -\sin\beta_1  \\
    \sin\beta_1 \sin\beta_2 & -\cos\beta_2 & \cos\beta_1\sin\beta_2
     \\
\end{array}
\right)\,.
\label{e:o}
\end{equation}

We now turn our attention to the physical scalar spectrum of the model. Since we are considering a potential with explicit CP conservation and a vacuum which does not spontaneously break CP, the neutral scalars have definite CP quantum numbers. The scalar spectrum of the model is composed of a pair of pseudoscalars, a trio of CP-even scalars and a pair of charged scalars, to be discussed in what follows.

In this work, we have studied the properties of the Higgs sector in the so-called Higgs alignment limit such that one of the physical
scalars coincides with the SM Higgs boson (i.e. features its mass and interactions). In order to ensure this in the input data prepared for
our parameter scans we would like to utilise an inversion procedure and require
the alignment limit at the level of input parameters. Such an inversion procedure would enable us
to express the parameters of the scalar potential in terms of physical masses, VEVs and mixing angles.

The mass terms for the pseudoscalar sector can be straightforwardly extracted from the scalar potential --
they will correspond to the terms quadratic in the $z_k$ ($k=1,2,3$) fields, after one has replaced the
expression for the doublets of \cref{eq:doub} into the potential \eqref{e:pot} and \eqref{e:pot-soft}.
One obtains
\begin{equation}\label{e:A1}
V_{P}^{\rm mass} = \begin{pmatrix}
z_1 & z_2 & z_3
\end{pmatrix} \, \frac{{M}_P^2}{2} \, \begin{pmatrix}
z_1\\  z_2\\ z_3\\
\end{pmatrix} \,,
\end{equation}
where ${\cal M}_{P}^2$ is the $3\times3$ pseudoscalar mass matrix
which takes a block-diagonal form in the Higgs basis with the help
of the orthogonal matrix $\mathcal{O}_\beta$ defined in~\eqref{e:o},
\begin{equation}\label{e:mp2x2}
\begin{split}
{B}^2_{P} \equiv \mathcal{O}_\beta \cdot {\cal M}_{P}^2 \cdot \mathcal{O}_\beta^{T} =& \begin{pmatrix}
0 & 0 & 0 \\
0 & {({B}_P^2)}_{22}  & {({B}_P^2)}_{23} \\
0 & {({B}_P^2)}_{23} & {({B}_P^2)}_{33} \\
\end{pmatrix} \,,
\end{split}
\end{equation}
where
\begin{equation}
\label{eq:3HDMP_pseudo_intermedium}
\begin{split}
\left( B^2_P \right)_{22} = & -\dfrac{1}{v_1 v_3 v_{13}^2}\left[2 \lambda _{10} \Big(v_1^5 v_3 + 2 v_1^3 v_3^3 + v_1 v_3^5 \Big) + \mu_{13}^2 v_{13}^4 + v_2 v_3^3 \mu_{21}^2 + v_1^3 v_2 \mu_{23}^2 \right]) \, , \\
\\
\left( B^2_P \right)_{32} = & -\dfrac{v}{v_{13}^2} \Big( v_1 \mu_{23}^2 - v_3 \mu_{21}^2 \Big)\, , \\
\\
\left( B^2_P \right)_{33} = & -\dfrac{v^2}{v_2 v_{13}^2} \Big( v_1 \mu_{21}^2 + v_3 \mu_{23}^2 \Big) \,.
\end{split}
\end{equation}

Thus, apart from the three VEVs, only $\lambda_{10}$ and the soft mass parameters, $\mu_{12}^2,\,\mu_{13}^2,\,\mu_{23}^2$, enter the pseudoscalar mass matrix. The line and column of zeroes in this matrix obviously tells us that it has a zero eigenvalue -- which of course is the neutral Goldstone boson responsible for the longitudinal polarisation of the massive $Z$ boson.

The pseudoscalar mass matrix is further diagonalized to the mass basis via an additional rotation,
\be
{\mathcal{O}}_{\gamma_1} \cdot {B}^2_{P} \cdot {\cal O}_{\gamma_1}^T = \left(\begin{array}{ccc}
	0 & 0 & 0 \\
	0 & m_{A1}^2 & 0 \\
	0 & 0 & m_{A2}^2 \\
\end{array}\right)\,, \label{e.Ogamma2}
\ee
where the matrix ${\cal O}_{\gamma_1}$ is defined as
\be
{\cal O}_{\gamma_1} =
\begin{pmatrix}
	1 & 0 & 0 \\
	0 & \cos\gamma_1 & -\sin\gamma_1 \\
	0 & \sin\gamma_1 & \cos\gamma_1 \end{pmatrix}   \,.
\ee
The full diagonalization of $M^2_P$ from \cref{e:A1} is therefore accomplished with the matrix product $\mathcal{O}_\beta \cdot \mathcal{O}_{\gamma_1}$, and this is important when one wishes to write the Yukawa interactions
between the two physical pseudoscalars and the physical quarks.

An analogous ~procedure can be performed in other sectors. For instance, the $3\times3$ charged Higgs sector mass matrix, ${M}_{C}^2$,
can also be block diagonalized by the same orthogonal matrix as:
\begin{eqnarray}
{B}^2_{C} \equiv \mathcal{O}_\beta \cdot {\cal M}_{C}^2 \cdot \mathcal{O}_\beta^T &=& \begin{pmatrix}
0 & 0 & 0 \\
0 & {({B}_C^2)}_{22}  & {({B}_C^2)}_{23} \\
0 & {({B}_C^2)}_{23} & {({B}_C^2)}_{33} \\
\end{pmatrix} \,. \label{e:BC2}
\end{eqnarray}
where again the single line and column of zeros yields a zero eigenvalue, the massless charged
Goldstone boson providing the longitudinal polarisation of the massive $W$ boson.
In the above matrix, we find
\begin{equation}
\begin{split}
{({B}_C^2})_{22} =& -\frac{1}{2 v_1 v_3 v_{13}^2} \Bigg[v_1^5 v_3 (2 \lambda_{10} + \lambda_8) + v_1 \Big( v_2^2 v_3^3 \lambda_7 +v_3^5 (2 \lambda_{10} +\lambda_8) \Big) + 2 v_1^4 \mu_{13}^2 + 4 v_1^2 v_3^2 \mu_{13}^2
\\ &
+ 2 v_3^3(v_3 \mu_{13}^2 + v_2 \mu_{21}^2) + v_1^3 \Big( 2 v_3^3 (2 \lambda_{10} + \lambda_8) + v_2^2 v_3 \lambda_9 + 2 v_2 \mu_{23}^2 \Big)\Bigg] \,, \\
{({B}_C^2)}_{23}  =& -\frac{v \left[v_1 v_2 v_3 (\lambda_7-\lambda_9)+2 v_3 \mu_{21}^2 - 2 v_1 \mu_{23}^2 \right]}{2 v_{13}^2} \,, \\
{({B}_C^2)}_{33} =&
-\frac{v^2 \left[ v_1^2 v_2 \lambda_{7} + 2 v_1 \mu_{21}^2 + v_3 (v_2 v_3 \lambda_9 + 2 \mu_{23}^2) \right]}{2 v_3 v_{13}^2} \,.
\end{split}
\end{equation}
Then, one switches to the mass basis in the charged scalar mass matrix as follows,
\begin{subequations}\label{e:BCrot}
	\begin{eqnarray}
	{\cal O}_{\gamma_2} \cdot ({B}_{C})^2 \cdot {\cal O}_{\gamma_2}^T = \left(\begin{array}{ccc}
	0 & 0 & 0 \\
	0 & m_{C1}^2 & 0 \\
	0 & 0 & m_{C2}^2 \\
	\end{array}\right)\,,
	\end{eqnarray}
	with the charged mixing matrix
	\begin{eqnarray}
	{\cal O}_{\gamma_2} =
	\begin{pmatrix}
	1 & 0 & 0 \\
	0 & \cos\gamma_2 & -\sin\gamma_2 \\
	0 & \sin\gamma_2 & \cos\gamma_2 \end{pmatrix}
	\label{e:Ogamma2} \,,
	\end{eqnarray}
\end{subequations}
and where $m_{C1}$ and $m_{C2}$ denote the masses of the two physical charged scalars,
$H_1^{\pm}$ and $H_2^{\pm}$, respectively.

Repeating the procedure above also for the CP-even states, we obtain
\begin{eqnarray}\label{e:neutralmassmat}
V_{S}^{\rm mass} =\begin{pmatrix}
h_1 & h_2 & h_3\\
\end{pmatrix} \frac{{M}_S^2}{2} \begin{pmatrix}
h_1\\  h_2\\ h_3\\
\end{pmatrix} \,,
\end{eqnarray}
where ${M}_S^2$ is a $3\times3$ symmetric mass matrix. In explicit form,
\begin{equation}
M_S^2 =
\scalebox{0.95}{$
	\begin{pmatrix}
	-\dfrac{\mu_{21}^2 v_2 + \mu_{13}^2 v_3 -2 \lambda_1 v_1^3 }{v_1} & v_1  v_2 (\lambda_4+\lambda_7)+\mu_{21}^2 & v_1
	v_3 (2 \lambda_{10} + \lambda_5 + \lambda_8) + \mu_{13}^2 \\
	v_1 v_2 ( \lambda_4 +\lambda_7) + \mu_{21}^2 & -\dfrac{\mu_{21}^2 v_1 +\mu_{23}^2 v_3 -2 \lambda_2 v_2^3}{v_2} & v_2
	v_3 (\lambda_6+\lambda_9)+\mu_{23}^2 \\
	v_1 v_3 (2 \lambda_{10}+\lambda_5+\lambda_8)+\mu_{13}^2 & v_2 v_3 (\lambda_6+\lambda_9)+\mu_{23}^2 & -\dfrac{\mu_{13}^2 v_1+\mu_{23}^2 v_2 -2 \lambda_3 v_3^3}{v_3} \\
	\end{pmatrix}$} \,. 	
\label{e:mselement}
\end{equation}
Switching to the Higgs basis,
\begin{equation}
\label{e:Ov}
\begin{pmatrix} H_0 \\ H'_1 \\ H'_2 \end{pmatrix} = \mathcal{O}_\beta \cdot
\begin{pmatrix} h_1 \\ h_2 \\ h_3 \end{pmatrix} \,.
\end{equation}
we notice that the state $H_0$ has the same gauge and Yukawa couplings at tree level as those of the SM Higgs boson.

The physical CP even scalars, $h$, $H_1$ and $H_2$, are obtained via
a different orthogonal rotation:
\begin{equation}
\label{e:Oalp}
\begin{pmatrix} h \\ H_1 \\ H_2 \end{pmatrix} = {\cal O}_\alpha
\begin{pmatrix} h_1 \\ h_2 \\ h_3 \end{pmatrix} \,,
\end{equation}
where ${\cal O}_\alpha$ is a $3\times 3$ orthogonal matrix which can be
conveniently parameterized as
\begin{subequations}
	\label{e:Oa}
	\begin{eqnarray}
	{\cal O}_\alpha &=& {R}_3 \cdot  {R}_1\cdot {R}_2 \,,
	\end{eqnarray}
	with,
	\begin{eqnarray}
	\label{e:R}
	{R}_1 = \left(\begin{array}{ccc}
	\cos \alpha_1 & \sin \alpha_1 & 0 \\
	-\sin \alpha_1 & \cos \alpha_1 & 0 \\
	0 & 0 & 1 \\
	\end{array}\right), \quad {R}_2 = \left(\begin{array}{ccc}
	\cos \alpha_2 & 0 & \sin \alpha_2  \\
	0 & 1 & 0 \\
	-\sin \alpha_2 & 0 & \cos \alpha_2 \\
	\end{array}\right),  \quad
	{R}_3 = \left(\begin{array}{ccc}
	1 & 0 & 0 \\
	0 & \cos \alpha_3 &  \sin \alpha_3  \\
	0 & -\sin \alpha_3 & \cos \alpha_3 \\
	\end{array}\right).
	\end{eqnarray}
\end{subequations}
Therefore, ${M}_S^2$ should be diagonalized via the following orthogonal transformation:
\begin{equation}\label{e:msdiag}
{\cal O}_\alpha \cdot {M}_{S}^2 \cdot {\cal O}_\alpha^T \equiv \begin{pmatrix}
m_h^2 & 0 & 0 \\
0& m_{H1}^2 & 0 \\
0 & 0 & m_{H2}^2 \\
\end{pmatrix}  \,,
\end{equation}
where the first eigenvalue corresponds to the SM-like Higgs boson
state $h$, with $m_h\simeq 125$ GeV.

Last but not least, it is instructive to discuss the alignment condition in our model. While a rotation to the Higgs basis is performed with the $\mathcal{O}_\beta$ matrix such that \cref{e:Ov} holds, the mass eigenstates can be written in terms of the Higgs basis ones as
\begin{equation}
\label{e:Oab}
\begin{pmatrix} h \\ H_1 \\ H_2 \end{pmatrix} = {\cal O}_\alpha {\cal O}_{\beta}^{\top}
\begin{pmatrix} H_0 \\ H_1^\prime \\ H_2^\prime \end{pmatrix} \,,
\end{equation}
where one defines the matrix
\begin{equation}\label{e:O}
	\mathcal{O} \equiv \mathcal{O}_\alpha \mathcal{O}_\beta^\top\,.
\end{equation}
The alignment limit is achieved once the SM-like Higgs boson completely overlaps with $H_0$ which, in practice, results in the condition
\begin{equation}\label{e:align}
	\mathcal{O}_{11} = 1\,.
\end{equation}
With our VEV parametrization in \cref{Eq:VeVs_Sines/Cosines} the alignment condition can be cast as
\begin{equation}\label{e:align-2}
	\cos \alpha_1 \cos \beta_2 \sin(\alpha_2 + \beta_1) + \sin \alpha_1 \sin \beta_2 = 1
\end{equation}
with reduces to the result in \cite{Das:2019yad} if one identifies
\begin{equation}
	\alpha_2 \leftrightarrow \alpha_1 \qquad \text{and} \qquad \beta_1 \to - \beta_1 + \dfrac{\pi}{2}\,.
\end{equation}
This means that putting $m_h=125$~GeV, $\alpha_1=\beta_2$ and $\alpha_2=-\beta_1 + \pi/2$ will ensure the presence of a 125~GeV SM Higgs boson in the spectrum -- that is the {\em exact} alignment
limit of this model, forcing the interactions between $h$ and the electroweak gauge bosons $Z$, $W$ (as well as with SM fermions, see below) to be exactly identical to those of the SM.

In practice, the exact alignment implies that $(\tilde{M}_S^2)_{11} = m_h^2$ and $(\tilde{M}_S^2)_{12} = (\tilde{M}_S^2)_{13} = 0$ where we define the Higgs basis mass matrix
\begin{equation}
	\tilde{M}_S^2 \equiv \mathcal{O}_\beta \cdot M_S^2 \cdot \mathcal{O}_\beta^\top\,.
\end{equation}
This can be further solved with respect to $\lambda_1$, $\lambda_2$ and $\lambda_{10}$ such that one can write
\begin{equation}\label{e:inversion}
	\begin{split}
	\lambda_1 =& \dfrac{1}{2 v_1^4} \Bigg[ m_h^2 (v_1^2 - v_3^2) + 2 v_3^4 \lambda_3 - v_1^2 v_2^2 (\lambda_4 + \lambda_7) + v_2^2 v_3^2 (\lambda_6 + \lambda_9) \Bigg] \,,
	\\
	\lambda_2 =& \dfrac{1}{2 v_2^2} \Big[ m_h^2 - v_1^2 (\lambda_4 + \lambda_7) - v_3^2(\lambda_6 + \lambda_9) \Big] \,,
	\\
	\lambda_{10} =& \dfrac{1}{2 v_1^2} \Big[ m_h^2 - 2 v_3^2 \lambda_3 - v_1^2 (\lambda_5 + \lambda_8) - v_2^2 (\lambda_6 + \lambda_9) \Big]\,.
	\end{split}
\end{equation}

At this point, it is instructive to summarise the above steps.
First of all, in order to make our numerical calculations technically feasible and time efficient, in this work the analysis of the scalar spectrum (couplings, mixing and masses) is performed entirely at tree level. We note that the scalar potential in Eqs.~\eqref{e:pot} and \eqref{e:pot-soft} contains sixteen real parameters. Among them, the quadratic parameters $\mu_{1}^2$, $\mu_{2}^2$ and $\mu_{3}^2$ can be traded in favor of the three VEVs, $v_1$, $v_2$ and $v_3$ or equivalently $v$, $\tan\beta_1$ and $\tan\beta_2$. In our numerical studies we take advantage of the exact alignment limit in order to randomly sample $\tan \beta_1$, $\tan \beta_2$, $\lambda_{3,\ldots,9}$ as well the soft parameters $\mu_{13}^2$, $\mu_{21}^2$ and $\mu_{23}^2$ such that, using \cref{e:inversion}, one obtains the correct $\lambda_1$, $\lambda_2$ and $\lambda_{10}$ quartic couplings compatible with an exact alignment of the SM-like Higgs boson. While off-alignment deviations are beyond the scope of this article, we provide in \cref{app:gen} generic formulas to obtain the gauge eigenbasis parameters if the physical masses and mixing angles are provided as inputs.

\subsection{The Yukawa sector}
\label{sec:Yukawa}

Alongside the scalar field transformations of \cref{e:stran}
the following quark fields are assumed to transform nontrivially
under the ${\rm U}(1)\times \mathbb{Z}_2$ flavor symmetry:
\begin{subequations}
\label{e:ftran}
    \begin{eqnarray}
    {\rm U}(1) &:& Q_{L3} \to e^{i\alpha} Q_{L3} \,, \qquad
    p_{R3} \to e^{2 i\alpha} p_{R3} \,, \\
    \mathbb{Z}_2 &:& Q_{L3} \to - Q_{L3} \,, \qquad
    p_{R3} \to -p_{R3} \,, \qquad n_{R3} \to -n_{R3} \,,
    \end{eqnarray}
\end{subequations}
with $\alpha$ the same arbitrary phase of \cref{e:stran},
and the rest of the quark fields remain unaffected under said symmetry transformations. In \cref{e:ftran}, as before, $Q_{La}= (p_{La}, n_{La})^T$ denotes
the left-handed quark doublet of the $a$-th generation whereas $p_{Ra}$
and $n_{Ra}$ denote the $a$-th generation (unrotated) up (positively charged) and down (negatively charged) type quark
singlets respectively. Notice the similarity between these transformation
laws and those of the BGL model, \cref{eq:sym}.

The quark Yukawa Lagrangian for a 3HDM will then have the general form
\begin{eqnarray}
 \ml_Y = -\sum_{k=1}^{3} \left[\bar{Q}_{La} (\Gamma_k)_{ab}\,
 \phi_k \, n_{Rb} +\bar{Q}_{La} (\Delta_k)_{ab}\,
 \tilde{\phi}_k \, p_{Rb} +{\rm h.c.} \right] \,,
\end{eqnarray}
where as before $\Gamma_k$ and
$\Delta_k$ stand for the Yukawa matrices in the down and
up quark sectors respectively. Due to specific charge assignments given
for the Higgs doublets and quark fields under ${\rm U}(1)\times \mathbb{Z}_2$ these Yukawa matrices will have the
following textures:
\begin{eqnarray}
\label{e:textures}
\Gamma_1 = \begin{pmatrix}
0 & 0 & 0 \\ 0 & 0 & 0 \\ \times & \times & 0 \end{pmatrix}, ~~~
\Delta_1 = \begin{pmatrix}
0 & 0 & 0 \\ 0 & 0 & 0 \\ 0 & 0 & 0 \end{pmatrix}, ~~~
\Gamma_2, \Delta_2 = \begin{pmatrix}
\times & \times & 0 \\ \times & \times & 0 \\ 0 & 0 & 0 \end{pmatrix}, ~~~
\Gamma_3, \Delta_3 = \begin{pmatrix}
0 & 0 & 0 \\ 0 & 0 & 0 \\ 0 & 0 & \times \end{pmatrix}.
\end{eqnarray}
Therefore, the quark mass matrices that emerge from these Yukawa matrices
have the following structure:
\begin{equation}
\label{e:mumd}
M_p = \frac{1}{\sqrt{2}}\sum_{k=1}^{3}\Delta_k v_k = \begin{pmatrix}
\times & \times & 0 \\ \times & \times & 0 \\ 0 & 0 & \times \end{pmatrix}, ~~~
M_n = \frac{1}{\sqrt{2}} \sum_{k=1}^{3} \Gamma_k v_k = \begin{pmatrix}
\times & \times & 0 \\ \times & \times & 0 \\ \times & \times & \times \end{pmatrix}.
\end{equation}
We then rotate from the $p$ and $n$ fields to the physical quark states $u$ and $d$ via rotation matrices $V_L$, $V_R$, $U_L$ and $U_R$ identical to those of \cref{eq:qrot}. We thus obtain diagonal mass matrices as in \cref{e:dudd}, and the CKM matrix is, as before, given by $V = V^\dagger_L U_L$. Let us now analyse carefully the Yukawa couplings between the neutral scalar eigenstates and the physical quarks, with particular attention to any FCNC couplings
which may arise.

In the alignment limit, with $\alpha_1=\beta_1$ and
$\alpha_2=\beta_2$, the physical scalar $h$ completely overlaps with $H_0$. In that limit, the other physical scalars, $H_1$ and $H_2$, will, in general, be an orthogonal mixture of the intermediate states defined above, $H'_1$ and $H'_2$.

Now, the terms in the Yukawa Lagrangian pertaining to the interactions
between CP-even scalars and quarks are
\begin{equation}
\ml_Y^{\rm CP~even} = -\frac{1}{\sqrt{2}} \left[
\bar{n}_L \left(\sum_{k=1}^{3}\Gamma_k h_k\right) n_R +
\bar{p}_L \left(\sum_{k=1}^{3}\Delta_k h_k\right) p_R +{\rm h.c.} \right]
\,,
\label{e:CPY}
\end{equation}
from which, using the rotation matrix of \cref{e:Ov} to express the $h_k$ in terms of
  $H_0$, we can obtain
\begin{equation}
\begin{split}
\ml_Y^{H_0} =& -\frac{H_0}{v} \left[
\bar{n}_L \left(\frac{1}{\sqrt{2}}\sum_{k=1}^{3}\Gamma_k v_k\right) n_R +
\bar{p}_L \left(\frac{1}{\sqrt{2}}\sum_{k=1}^{3}\Delta_k v_k\right) p_R +{\rm h.c.} \right] \,, \\
=& -\frac{H_0}{v} \left[
\bar{d}_L D_d d_R +
\bar{u}_L D_u u_R +{\rm h.c.} \right] \,.
\end{split}
\label{e:H0}
\end{equation}
In writing the last step, we have made use of \cref{e:mumd,e:dudd}.
Thus we see that $H_0$ possesses SM like Yukawa coupling at tree
level. This is a close analogy to the BGL model, where we explained how,
in the exact 2HDM alignment limit, the $h$ state had identical Yukawa
interactions to those of the SM Higgs boson.

In a similar manner, we can write down the Yukawa couplings of $H'_1$
and $H'_2$ with the down type quarks as follows:
\begin{equation}
\ml_Y^{H'_1, H'_2} = -\frac{H'_1}{v} \bar{d}_L N_{d1} d_R
-\frac{H'_2}{v} \bar{d}_L N_{d2} d_R +{\rm h.c.} \,,
\end{equation}
where the matrices $N_{d1}$ and $N_{d2}$ are given by
\begin{subequations}
\label{e:nd}
    \begin{equation}
    \begin{split}
    N_{d1} =& \frac{v}{\sqrt{2} v_{13}} U_L^\dagger
    (\Gamma_1 v_3 - \Gamma_3 v_1) U_R \,, \\
    N_{d2} =& U_L^\dagger \left[\frac{v_2}{v_{13}} \frac{1}{\sqrt{2}}
    \left(\Gamma_1 v_1+\Gamma_3 v_3 \right) -\frac{v_{13}}{v_2} \frac{1}{\sqrt{2}} \Gamma_2 v_2 \right] U_R \,.
    \end{split}
    \end{equation}
\end{subequations}
To simplify further the expressions for $N_{d1}$ and $N_{d2}$, we
go back to the textures of the mass matrices in \Eqn{e:mumd}. From
the block diagonal structure of $M_u$, one can conclude that the
corresponding bidiagonalizing matrices, $V_L$ and $V_R$, should
have block diagonal structures too. In fact, we can choose
\begin{equation}
V_L =
\begin{pmatrix}
\times & \times & 0 \\ \times & \times & 0 \\ 0 & 0 & 1 \end{pmatrix}
\end{equation}
with the understanding that the phase of $(M_u)_{33}$ can always
be absorbed into $(V_R)_{33}$. Here, unlike the BGL example of
section~\ref{sec:bgl}, we are choosing to single out the third family.
Then, from \Eqn{e:CKM} we obtain
\begin{equation}
\label{e:V3A}
  (U_L)_{3A} = V_{3A} \,,
\end{equation}
which means that the third row of $U_L$ is identical to that of
the CKM matrix, as occurred in the 2HDM BGL case. To proceed further, it is useful to define the
following projection matrix
\begin{equation}
P = \begin{pmatrix}
0 & 0 & 0 \\ 0 & 0 & 0 \\ 0 & 0 & 1 \end{pmatrix} \,.
\end{equation}
Thus, in view of the structures of the Yukawa matrices, we
obtain the following relations in the down quark sector:
\begin{eqnarray}
\label{e:Peq}
\Gamma_3 = (\Gamma_3)_{33} P \,, \qquad
\frac{1}{\sqrt{2}} \left(\Gamma_1 v_1+\Gamma_3 v_3 \right) =
P\,M_d \,.
\end{eqnarray}
Using \Eqs{e:V3A}{e:Peq}, the expressions for $N_{d1}$ and $N_{d2}$
can now be simplified so that:
\begin{subequations}
    \label{e:ndf}
    \begin{eqnarray}
    (N_{d1})_{AB} &=& \frac{v\, v_3}{v_1 v_{13}} V^*_{3A} V_{3B}
    (D_d)_{BB} -\frac{1}{\sqrt{2}} \frac{v\, v_{13}}{v_1}
     (\Gamma_3)_{33} V^*_{3A} (U_R)_{3B}  \,,
     \label{eq:nd1} \\
    (N_{d2})_{AB} &=& \frac{v_{13}}{v_2} (D_d)_{BB}\delta_{AB}
    +\left(\frac{v_{13}}{v_2}+\frac{v_2}{v_{13}}\right)
     V^*_{3A} V_{3B} (D_d)_{BB}  \,.
    \end{eqnarray}
\end{subequations}
These equations tell us that the FCNC interactions of $H^\prime_2$ are {\em exactly}
BGL-like -- all off-diagonal elements in $N_{d2}$ are CKM-suppressed. That however
is not the case for $H^\prime_1$ -- the first term in the right-hand side of Eq.~\eqref{eq:nd1}
is a matrix whose off-diagonal entries are suppressed by {\em two} CKM matrix elements, as in
the BGL model, but the second term's FCNC couplings are suppressed by only {\em one}
CKM matrix element.
To estimate the size of $(U_R)_{3B}$, we note that if $(\Gamma_1)_{31}$
and $(\Gamma_1)_{32}$ were zero in \Eqn{e:textures}, then we could choose
$U_R$ to be block diagonal as well. However, in view of the smallness
of the off-diagonal elements in the CKM matrix, the elements of
$\Gamma_1$ should also be quite small. Therefore it is reasonable
to assume that the elements $(U_R)_{3B}$ (for $B\ne 3$) are also
small. Considering all these facts, one can expect that the FCNC
couplings in the down quark sector controlled by $N_{d1}$ and $N_{d2}$
will be under control.

In this work, as was stated above, we work in the exact alignment
limit implying that the physical state $h$ couples to the SM fermions and
gauge bosons with the same strength as the SM Higgs boson.
Ultimately, of course, we will be close to the alignment limit -- the current LHC data requires it, see \cref{sec:exp} -- but not necessarily
{\em exactly} in it, so the physical CP-even/odd scalar states
would have FCNC interactions given by linear combinations
of the $N_{d1}$ and $N_{d2}$ matrices via the rotation matrices defined
in \cref{sec:sca}. As such, one obtains a model
that is not as ``clean'' as the 2HDM BGL, but where one still sees how
FCNC interactions arise which are CKM-suppressed due to the
symmetries imposed upon the potential -- the suppression is therefore
natural and not the result of a fine tuning. A quantitative study of
the possible impact arising from off-alignment corrections on FCNC observables in the current model is a subject of a dedicated work in the future.

A similar exercise in the up quark sector would reveal that there
are no scalar mediated FCNC at tree level in the up
sector. This is due to the special structures of the up type
Yukawa matrices, which, in turn, are dictated by the fermionic
charges given in \Eqn{e:ftran}. But it should be noted that, just
like in the usual BGL models, the charges in \Eqn{e:ftran} can be
appropriately shuffled so that the tree level FCNC couplings reside
entirely in the up quark sector instead of the down quark sector.
And within each sector one still has the possibility of choosing
FCNC associated with a given family.
However, we choose the current variant -- FCNC in the down sector,
associated with the third family -- because it will be the
most restrictive one. There is a wealth of experimental data limiting such FCNCs,
thus this version of the model will be the most restricted one. Therefore,
if this version can survive all constraints we will throw at it,
other versions would more easily pass the same constraints
and may be the subject of future studies.

In passing, it should be mentioned that the leptonic fields are
assumed to couple only to $\phi_1$ in the Yukawa sector. This
can be achieved very simply by assigning the following
transformations to the leptonic fields
\begin{subequations}
    \label{e:ltran}
    \begin{eqnarray}
    {\rm U} &:& L_{La} \to e^{i\alpha} L_{La} \,, \qquad
    \ell_{Ra} \to \ell_{Ra} \,, \\
    \mathbb{Z}_2 &:& L_{La} \to - L_{La} \,, \qquad
    \ell_{Ra} \to \ell_{Ra} \,,
    \end{eqnarray}
\end{subequations}
where $L_{La}= (\nu_{La}, \ell_{La})^T$ and $\ell_{Ra}$ denote,
respectively, the left-handed lepton doublet and the right handed
charged lepton singlet of $a$-th generation. Since the charged
leptons receive their masses from a single scalar doublet, there
will be no FCNC couplings at tree level in the leptonic sector.
Thus, all constraints from observables such as $\mu\to e \gamma$ are
automatically satisfied, since for simplicity in our model the lepton number is assumed to be exactly conserved. Additionally, it is also worth mentioning that in this first study of a 3HDM BGL-like scenario we do not
introduce right handed neutrinos, {\it i.e.}, neutrinos are
assumed to be massless in our model.

One major challenge in producing a BSM theory with a non-trivial
Yukawa sector, {\em i.e.} with FCNC interactions, resides in being able to
successfully fit the quark mass spectrum and CKM mixing, simultaneously. In fact, it is a highly non-trivial -- and time-consuming -- task to find values for Yukawa couplings and scalar VEVs capable of fitting quark masses which differ by many orders of magnitude. Add to that the difficulty in having to simultaneously being
capable of fitting the CKM matrix entries and a simultaneous fit to the quark and scalar sector
becomes a very difficult achievement.

In this work, we follow an {\em inversion
procedure} in some ways similar to that implemented in \cref{app:gen} for the scalar sector. In essence, the inversion here means expressing the Lagrangian parameters of a given BSM scenario (partially or completely) in terms of physical observables and mixing angles connecting the gauge basis with the physical one. In the model under consideration, such inversion can be {\it unambiguously} realized since the corresponding system of equations is linear and non-singular and thus yields a unique non-trivial solution, upon an appropriate choice of input parameters.

The inversion procedure in the Yukawa sector consists in, literally,
inverting the usual fitting logic of the Yukawa sector: instead of scanning over Yukawa couplings and VEVs defining some sort of $\chi^2$ function whose minimization would yield an acceptable quark mass spectrum and CKM matrix, we do the opposite.
Namely, quark masses and CKM matrix elements are our initial inputs, and we scan over the {\em bidiagonalization matrices} which pass from the interaction basis to the mass eigenstate basis.

To make this clear, let us begin with the diagonal quark mass matrices,
$D_u = diag(m_u\,,\,m_c\,,\,m_t)$ and $D_d = diag(m_d\,,\,m_s\,,\,m_b)$.
Within the considered BGL-like 3HDM, they are the result of the bidiagonalization of the
(interaction basis) matrices of Eq.~\eqref{e:mumd}, via $3\times 3$ unitary matrices
$U_L$, $U_R$, $V_L$ and $V_R$, such that
\be
D_u = V^\dagger_L \,M_p\,V_R\;\;\; , \;\;\; D_d = U^\dagger_L \,M_n\,U_R\, ,
\ee
and the CKM matrix $V$ defined above.
Using the unitarity of the rotation matrices we can invert the above equations, and since the definition of the CKM matrix implies that $U_L = V_L \,V$, we can write
\begin{equation}
	\begin{split}
	M_p =& V_L\,D_u\,V^\dagger_R \,=\,\begin{pmatrix}
	(\Delta_2)_{11} v_2 & (\Delta_2)_{12} v_2 & 0 \\
	(\Delta_2)_{21} v_2 & (\Delta_2)_{22} v_2 & 0 \\ 0 & 0 & (\Delta_2)_{33} v_3 \end{pmatrix} \,, \\
	M_n =& V_L \,V\,D_d\,U^\dagger_R\,=\, \begin{pmatrix}
	(\Gamma_2)_{11} v_2 & (\Gamma_2)_{12} v_2 & 0 \\
	(\Gamma_2)_{21} v_2 & (\Gamma_2)_{22} v_2 & 0 \\
	(\Gamma_1)_{31} v_1 & (\Gamma_1)_{32} v_1 & (\Gamma_3)_{33} v_3 \end{pmatrix} \,,
	\end{split}
	\label{eq:inv}
\end{equation}
where $U_L$ has been replaced by $V_L$ and the CKM matrix.
In \cref{eq:inv} the quark masses, CKM matrix and VEVs will be the inputs (the VEVs obtained from a previous partial scan of the scalar sector, already ensuring that the alignment limit is satisfied).
The unknowns are the rotation matrices $V_L$, $V_R$ and $U_R$.
Since any $U(3)$ matrix can be parameterized by three angles and six phases we have a total of 27 free parameters
with which one would attempt a fit to \cref{eq:inv}. However, using our inversion method there is no fit involved and, due to the Yukawa textures, $V_{L,R}$ become $2\times 2$ block diagonal matrices. This means that, in the up sector, we can parameterize our rotation to the mass basis with two angles, $\alpha_L$ and $\alpha_R$, defined as
\begin{equation}
	V_L  \,=\,\begin{pmatrix}
	\cos \alpha_L & \sin \alpha_L & 0 \\
	-\sin \alpha_L & \cos \alpha_L & 0 \\
	0 & 0 & 1 \end{pmatrix} \,,
	\qquad
	V_R \,=\, \begin{pmatrix}
	\cos \alpha_R & \sin \alpha_R & 0 \\
	-\sin \alpha_R & \cos \alpha_R & 0 \\
	0 & 0 & 1 \end{pmatrix} \,.
	\label{eq:VLVR}
\end{equation}
Therefore, the five {\em real} non-zero elements of $\Delta$ in $M_p$ are traded for $\alpha_L$, $\alpha_R$ and the physical up-type quark masses $m_u$, $m_c$ and $m_t$. In the down sector, on the other hand, both $U_{L}$ and $U_{R}$ are generic $3\times 3$ matrices. While the former is fixed as  $U_{L} = V_L V$, with the CKM matrix expressed in the Wolfenstein form ( see PDG for details \cite{Zyla:2020zbs}), the latter can be parameterized with three angles $\beta_R^{1,2,3}$ if we assume, as we do in our numerical implementation, that the only complex CP-phase comes from the CKM matrix. This means that, for our scenario, the seven non-zero elements of $\Gamma$ in $M_n$ can be consistently described with eight real couplings, which are replaced by the four Wolfenstein parameters $\lambda$, $A$, $\bar{\rho}$ and $\bar{\eta}$, three quark masses, $m_d$, $m_s$ and $m_b$ and a randomly generated angle which we call $\beta_R$. In particular, denoting the solutions of \cref{eq:inv} for the angles as $\beta_R^2 = \beta_R^2 (\lambda,A,\bar{\rho}, \bar{\eta}, m_b, m_s, m_d, \beta_R)$ and $\beta_R^3 = \beta_R^3 (\lambda,A,\bar{\rho}, \bar{\eta}, m_b, m_s, m_d, \beta_R)$, which are numerically computed, one can express $U_R$ as
\begin{equation}
	U_R \,=\, \begin{pmatrix}
	\cos \beta_R \cos \beta_R^2 & \sin \beta_R \cos \beta_R^2 & \sin \beta_R^2 \\
	-\cos \beta_R \sin \beta_R^2 \sin \beta_R^3 -\sin \beta_R \cos \beta_R^3 & \cos \beta_R \cos \beta_R^3 - \sin \beta_R \sin \beta_R^2 \sin \beta_R^3 & \cos \beta_R^2 \sin \beta_R^3 \\
	- \cos \beta_R \sin \beta_R^2 \cos \beta_R^3 + \sin \beta_R \sin \beta_R^3 & - \cos \beta_R \sin \beta_R^3 -\sin \beta_R \sin \beta_R^2 \cos \beta_R^3 & \cos \beta_R^2 \cos \beta_R^3 \end{pmatrix} \,,
	\label{eq:UR}
\end{equation}
with $\beta_R$ a free parameter.

All in all, given the physical quark masses and CKM mixing (with a complex CP-phase), our numerical sampling of the Yukawa sector is consistently achieved solely with three angles, $\alpha_L$, $\alpha_R$ and $\beta_R$, upon inversion of \cref{eq:inv}. Therefore with the VEVs and the values obtained for the nonzero entries, it is a simple matter to reconstruct the elements of the $\Gamma$ and $\Delta$ matrices -- notice that each non-zero entry of $M_p$ ($M_n$) has the contribution of a
single $\Delta$ ($\Gamma$) matrix element such that the reconstruction of the Yukawa matrices is unequivocal.

With this simple procedure we ensure that the very different quark masses are automatically and exactly reproduced for every single sampled point, and so does the non-trivial structure of the CKM matrix. The computational challenge then is to scan over three rotation angles in order to reproduce the correct flavour observables in consistency with strict experimental constraints.

\section{Constraints on the model}
\label{sec:cons}

Any realistic BSM theory needs to do, at least, as good a job as the SM in describing well measured particle physics properties. For multi-scalar scenarios, there is a wealth of theoretical and experimental bounds that are imposed upon the model's parameter space. The main challenge then is to find the domains of validity for a given model for which the parameter space points pass all the relevant restrictions. In this section, we summarize the main theoretical and experimental constraints that need to be satisfied in order to validate our BGL-like 3HDM framework.

\subsection{Theoretical constraints}
\label{sec:theo}

For models with a scalar content larger than that of the SM, special attention needs to be focused on the possibility of the scalar potential becoming unbounded from below i.e.~tending to minus infinity for some direction along which the fields are assuming arbitrarily large values. This imposes constraints on the model's scalar quartic couplings, as the
quartic part of the potential clearly dominates over the quadratic (or
even an eventual cubic) one when the scalar fields tend to infinity.
This is already a concern for the SM -- it is the reason why the
SM Higgs quartic coupling $\lambda$ is taken positive. For the 2HDM, generic conditions were found in Refs.~\cite{Ivanov:2006yq,Ivanov:2007de} but for the 3HDM, on the other hand, there is no such generic boundedness-from-below prescription that can be straightforwardly implemented in a parameter scan (see \cite{Faro:2019vcd} for a $\U{} \times \U{}$ version of the 3HDM). Still, some {\em necessary} conditions are easy to find. Analysing the shape of the scalar potential of Eq.~\eqref{e:pot}, we see that if one takes each of the doublets $\phi_i$ to infinity separately, the potential will tend to $-\infty$ unless
\be
\lambda_1 \,>\,0\;\;\; , \;\;\; \lambda_2 \,>\,0\;\;\; , \;\;\; \lambda_3 \,>\,0\,.
\ee
Likewise, following a procedure similar to the one used in the 2HDM~\cite{Branco:1996bq}, if one
takes two doublets $(i,j)$ to infinity but such that $\phi^\dagger_i \phi_j = 0$ (which is realised, for instance, when
the upper component of one of the doublets and the lower component of the other doublet are zero) one obtains a positive value of the potential for any value of the fields if
\be
\lambda_4\,>-\,2\sqrt{\lambda_1 \lambda_2} \;\;\; , \;\;\;
\lambda_5\,>-\,2\sqrt{\lambda_1 \lambda_3} \;\;\; , \;\;\;
\lambda_6\,>-\,2\sqrt{\lambda_2 \lambda_3} \,.
\ee
We can also adapt the boundedness-from-below necessary conditions from Ref.~\cite{Moretti:2015cwa} (see Eqs.~(21)--(24) there), being careful with the fact that the potential of that work is different from the one considered here (our potential has a more restrictive symmetry, and therefore has fewer quartic couplings). This translates into a generalisation of the above conditions, which become
\be
\lambda_4\,>-\,2\sqrt{\lambda_1 \lambda_2}\,-\,{\rm min}(0,\lambda_7) \;\; , \;\;
\lambda_5\,>-\,2\sqrt{\lambda_1 \lambda_3} \,-\,{\rm min}(0,\lambda_8 - 2|\lambda_{10}|)\;\; , \;\;
\lambda_6\,>-\,2\sqrt{\lambda_2 \lambda_3} \,-\,{\rm min}(0,\lambda_9) \,.
\ee
Ultimately, these necessary conditions eliminate a great deal of parameter space. Even though they are not the {\em sufficient}
ones, they are still expected to cover most of the parameter space regions that lead to a bounded-from-below potential.

Another strong constraint on the quartic couplings of potential is the requirement for the theory to be unitary. For the SM, this implies an upper bound on the mass of the Higgs boson~\cite{Lee:1977yc,Lee:1977eg},
and similar studies have been applied to the 2HDM (general unitarity conditions are presented in Ref.~\cite{Ginzburg:2005dt}) and other models with a larger scalar content. Essentially, the method consists in computing all scalar-scalar $J = 0$ scattering amplitudes (usually denoted $a_0$) and requiring that they respect unitarity in the high energy regime. This translates into an upper bound on those amplitudes, $|a_0| < 1/2$. Theories with many scalars complicate the calculation somewhat due to a growing multiplicity of such scattering amplitudes and bounds must be imposed upon eigenvalues of the S-matrix. General unitarity bounds for a 3HDM with a $\mathbb{Z}_2\times \mathbb{Z}_2$-symmetric potential are presented in Ref.~\cite{Moretti:2015cwa},
of which our ${\rm U}(1)\times \mathbb{Z}_2$ symmetry is a special case. Since our model has a larger symmetry, it has less parameters than that of Ref.~\cite{Moretti:2015cwa} and one could read off from Eqs.~(91) to (102) of that article the unitarity constraints imposed upon the quartic couplings. However, as explained below, we will use the \texttt{SARAH/SPheno} machinery instead to take into account these bounds.

Finally, a ``standard" constraint on multiscalar models is to verify their compliance with electroweak precision bounds i.e.~constraints on the oblique $S,T$ and $U$ parameters. Models with $N$ Higgs doublets automatically satisfy $\rho = 1$ at tree-level, which means that bounds on the oblique parameter $S$ will be easily satisfied. However, that is no longer true for the $T$ parameter, which typically needs to be computed for any given model. Constraints on $T$ typically enforce, for very high masses, degeneracies between the extra scalars in the mass spectrum of the model. The results from Ref.~\cite{Moretti:2015cwa} are of no help to us in this case, as the expressions for the oblique
parameters given there are only valid for a 3HDM version of the inert type (where one of the doublets is
VEVless and naturally decouples from the gauge sector).

Instead of computing the unitarity bounds and oblique corrections analytically, in this work, we have
adopted another strategy and used the publicly available \texttt{SARAH/SPheno} framework \cite{Porod:2003um,Porod:2011nf,Staub:2013tta} which enables one to
compute them numerically in a given particle physics model for a particular parameter space point (for an earlier implementation of this procedure, see e.g. Ref.~\cite{Morais:2019aqz}).
In our numerical analysis substantiated in detail below, we have implemented the 3HDM model under consideration into this framework where both the unitarity constraints and the electroweak precision bounds on $S,T,U$ parameters have been consistently incorporated in the parameter scans designed to search for the valid physical regions of the model. In particular, the oblique parameters are computed by \texttt{SPheno-4.0.4} at one loop-order using tree-level masses and then are verified against
the experimental bounds for every parameter space point.

\subsection{Experimental constraints}
\label{sec:exp}

Since the discovery of the Higgs-like state in 2012 the LHC collaborations
have been able to verify that its properties correspond to those expected for the SM Higgs boson, within rather small experimental error bars. In practical terms, this means that the couplings of the 125 GeV $h$ state to electroweak gauge bosons and fermions in a BSM model cannot deviate too much from the corresponding SM couplings. A convenient way of parameterizing the Higgs interaction strengths is by introducing the $\kappa$-formalism, defining the dimensionless quantities,
\be
\kappa_X^2\,=\,\frac{\Gamma^{\rm BSM}(h\to X)}{\Gamma^{\rm SM}(h\to X)} \,,
\ee
through the decay widths, $\Gamma$, of the Higgs boson to a certain final state $X$ (typically, $ZZ$, $WW$, $\tau\bar{\tau}$ and $b\bar{b}$), computed both in a given BSM scenario as well as in the SM. This definition implies that the exact SM behaviour would correspond to $\kappa = 1$. Notice that current LHC measurements~\cite{Aad:2015zhl,Khachatryan:2016vau} only allow deviations from unity roughly up to 20\% for the several final states.

Requiring $h$ to be SM-like naturally suppresses the couplings to gauge
bosons of the heavier CP-even states $H_1$ and $H_2$ (these three states'
gauge couplings obey a sum rule, due to gauge invariance). Since
one of the most constraining channels in the search for heavier resonances
at the LHC is precisely via di-$Z$ boson production, most of the constraints coming from those searches are automatically satisfied. However, there is still an ample parameter space for which the heavier scalars also have suppressed production cross sections and could thus have avoided detection so far.

The basic properties of the Higgs boson in the BGL-like 3HDM under consideration, such as its physical couplings, decay rates and branching ratios, have been computed numerically in \texttt{SPheno-4.0.4} for each input parameter-space point in a dedicated numerical scan. A set of input
points has been prepared in a separate special routine using an inversion procedure in the scalar sector in order to reproduce the exact Higgs alignment limit, with fixed VEV $v\simeq 246$ GeV and Higgs mass $m_h\simeq 125$ GeV values, as detailed above. Then, a candidate point has been chosen to satisfy the electroweak precision bounds, the boundedness-from-below and unitarity constraints discussed in the previous subsection. On top of that, in order to take into account the latest data from the LHC on the properties of the observed 125 GeV state we used \texttt{HiggsSignals-2.2.3beta}~\cite{Bechtle:2013xfa}, whereas limits from the LEP, Tevatron and LHC direct searches for new CP-even and CP-odd scalars were taken into account through the use of \texttt{HiggsBounds-5.3.2beta}~\cite{Bechtle:2008jh,Bechtle:2011sb,Bechtle:2013wla}.
The latter two packages are linked to \texttt{SPheno-4.0.4} such that they get all the necessary inputs containing masses and decay widths of the SM and BSM Higgs states for each parameter-space point. A point that passes all these constraints is considered to be valid, at least, from the point of view of the scalar sector constraints.

As for the flavour sector, there are numerous observables that have to be computed and checked for each parameter space point that satisfies the scalar sector constraints. The inverted procedure outlined in \cref{sec:Yukawa} already ensures that our choice of the parameter-space points, even before a numerical scan, automatically reproduces the correct measured values of the quark masses and of the CKM matrix elements. But the presence of charged scalars, as well as neutral ones
with tree-level CKM-suppressed FCNC interactions, means that we must verify particularly sensitive quantities, such as the $b\rightarrow s \gamma$ width, the neutral $K$- and $B$-meson mass differences and the CP-violating $\epsilon_K$ phase, among others. To this end, we employ the \texttt{FlavourKit} package \cite{Porod:2014xia} as part of the \texttt{SPheno} tool in order to compute the full list of Wilson coefficients for a given parameter-space point. The Wilson coefficients
are then passed over to the \texttt{Flavio} python package \cite{Straub:2018kue} enabling to compute all the relevant flavour physics observables for a given point and to compare them to the corresponding SM values.

In particular, out of an extensive list, we have found that the most relevant~\footnote{``Relevant" in the sense
that these observables are may have, in our calculations, the largest deviations from the SM-expected values, and thus harder to fit; but
when one finds a set of parameters which fits these five quantities, the remaining flavour observables are
found to be in general
agreement with SM expectations.} quark flavour violation (QFV) observables to consider are $B \rightarrow \chi_s \gamma$, $B_s \rightarrow \mu \mu$, $\Delta M_s$ , $\Delta M_d$ and $\varepsilon_K$, which we collectively denote as $\mathcal{O}_\mathrm{3HDM}$. In our analysis we quantify deviations from the SM prediction as a ratio $\mathcal{O}_\mathrm{3HDM} / \mathcal{O}_\mathrm{SM}$. Defining the QFV experimentally measured value, its experimental error and the SM prediction theoretical uncertainty as $\mathcal{O}_\mathrm{Exp}$, $\sigma_\mathrm{Exp}$ and $\sigma_\mathrm{SM}$ respectively, one can use the error propagation formalism to obtain the error associated to the ratio $\mathcal{O}_\mathrm{Exp} / \mathcal{O}_\mathrm{SM}$ that reads as
\begin{equation}\label{e:error}
\sigma = \dfrac{1}{\mathcal{O}_\mathrm{SM}} \sqrt{\sigma_\mathrm{Exp}^2 + \sigma_\mathrm{SM}^2 \dfrac{\mathcal{O}_\mathrm{Exp}^2}{\mathcal{O}_\mathrm{SM}^2}} \ .
\end{equation}
In \cref{tab:err} we numerically determine $\sigma$ for each of the five QFV observables mentioned above which will later be used to define the error bars in our plots.
\begin{table}[htb!]
	\centering
	\begin{tabular}{l|ccccc}
		Channel & $\mathcal{O}_\mathrm{SM}$         & $\sigma_\mathrm{SM}$ & $\mathcal{O}_\mathrm{Exp}$ & $\sigma_\mathrm{Exp}$ & $\sigma$  \\ \cline{1-6}
		$\mathcal{BR} ( B \rightarrow \chi_s \gamma )$ & $3.29 \times 10^{-4}$ & $1.87 \times 10^{-5}$ & $3.32 \times 10^{-4}$ & $0.16 \times 10^{-4}$ &  $0.075$ \\
		$\mathcal{BR} ( B_s \rightarrow \mu \mu ) $ & $3.66 \times 10^{-9}$ & $1.66 \times 10^{-10}$ & $2.80 \times 10^{-9}$ & $0.06 \times 10^{-9}$  &  $0.038$    \\
		$\Delta M_d$ (GeV) & $3.97 \times 10^{-13}$ & $5.07 \times 10^{-14}$ & $3.33 \times 10^{-13}$  &  $0.013 \times 10^{-13}$ & $0.11$  \\
		$\Delta M_s$ (GeV) & $1.24 \times 10^{-11}$ & $7.08 \times 10^{-13}$ & $1.17 \times 10^{-11}$ &  $0.0014 \times 10^{-11}$ & $0.054$ \\
		$\varepsilon_K $ & $1.81 \times 10^{-3}$  & $2.00 \times 10^{-4}$ & $2.23 \times 10^{-3}$  & $0.011 \times 10^{-3}$  & $0.14$
	\end{tabular}
	\caption{The measured QFV observables values and their respective uncertainties were taken from \cite{Tanabashi2018}. The SM prediction and the theoretical uncertainties are taken from \texttt{Flavio}. }
	\label{tab:err}
\end{table}

Given the BGL-like nature of the FCNC interactions in the considered 3HDM, the QFV observables appear to be already in the right ballpark of typical values being not too far from the SM results (see below). Thus, there is no need to include the flavour observables in the numerical fit -- we have just computed those observables for each point that has passed the theoretical, electroweak precision tests and Higgs sector constraints. In other words, when scanning over the BGL-3HDM parameter space requiring exact alignment and
using the inversion procedure described for the fermion sector, the parameters found already give QFV observable
values not too far from the expected values. As such there is no practical need to include fitting those variables
along with the scalar sector (a task which would be highly work-intensive), since the overall efficiency of the scan
is not compromised by points not complying with QFV observables. Let us now turn to a discussion of our numerical results.

\section{Results}
\label{sec:res}

As it was discussed above, we use an inversion procedure in order to automatically obtain the correct quark masses, CKM matrix elements, the measured Higgs boson mass as well as an exact alignment of the latter to the SM, for all sampled parameter space points. For such a purpose we have built an internal routine, that we shall denote as {\em pre-SPheno}, where \cref{e:inversion} and an expanded form of \cref{eq:inv} are implemented. With this we compute the gauge eigenbasis Lagrangian parameters and then proceed with the calculation passing them to \texttt{SPheno}. The input free parameters were then randomly sampled in the ranges given in \cref{tab:sample}.
\begin{table}[htb!]
	\centering
	\begin{tabular}{c|c|c|c}
		\toprule
		$|\mu_{21}^2|$, $|\mu_{13}^2|$, $|\mu_{23}^2|$ $(\mathrm{TeV}^2)$ & $|\lambda_{3,\ldots,9}|$ & $\beta_1, \beta_2$ & $\alpha_L$, $\alpha_R$, $\beta_R$  \\
		\midrule
		$  \left[ 0.1^2 , 10^2 \right]$ & $ \left[10^{-7}, 4 \pi\right]$ &  $\left[0,\dfrac{\pi}{2}\right[$ & $\left[0,2 \pi\right]$
		\\
		\bottomrule
	\end{tabular}
	\caption{Ranges used for the input parameters in our numerical scan. The values of the quark masses, of the CKM mixing elements and of the Higgs boson mass were fixed to their central values according to \cite{Zyla:2020zbs}.}
	\label{tab:sample}
\end{table}

In essence, for for all that follows, we have imposed {\em a priori} certain basic constraints implemented analytically: the correct electroweak symmetry breaking must occur, thus the doublets have VEVs satisfying $v_1^2 + v_2^2 + v_3^2 = (246$ ${\rm GeV})^2$;
the lightest CP-even mass is fixed to be 125 GeV; the $\lambda_1$, $\lambda_2$ and $\lambda_{10}$ quartic scalar couplings where chosen to provide the exact Higgs alignment limit; and the remaining angles in the fermion sector besides $\alpha_L$, $\alpha_R$ and $\beta_R$ were calculated to comply with the correct quark masses and CKM mixing. All the other constraints, such as the necessary boundedness-from-below conditions described in \cref{sec:theo}, the unitarity, electroweak
precision and Higgs-sector phenomenological constraints are checked numerically using the coupled chain
of computer packages -- \texttt{SPheno}, \texttt{HiggsSignals} (HS) and
\texttt{HiggsBounds} (HB) -- in a dedicated numerical scan, while the flavour physics
constraints are verified {\em a posteriori} using \texttt{FlavourKit} and \texttt{Flavio} tools.

Let us now investigate the effect that the various constraints, both on the scalar and flavour sectors, have on the allowed parameter space of the model.

\subsection{Allowed parameter space}

\begin{figure}[h!]
	\centering
	\includegraphics[width=\textwidth]{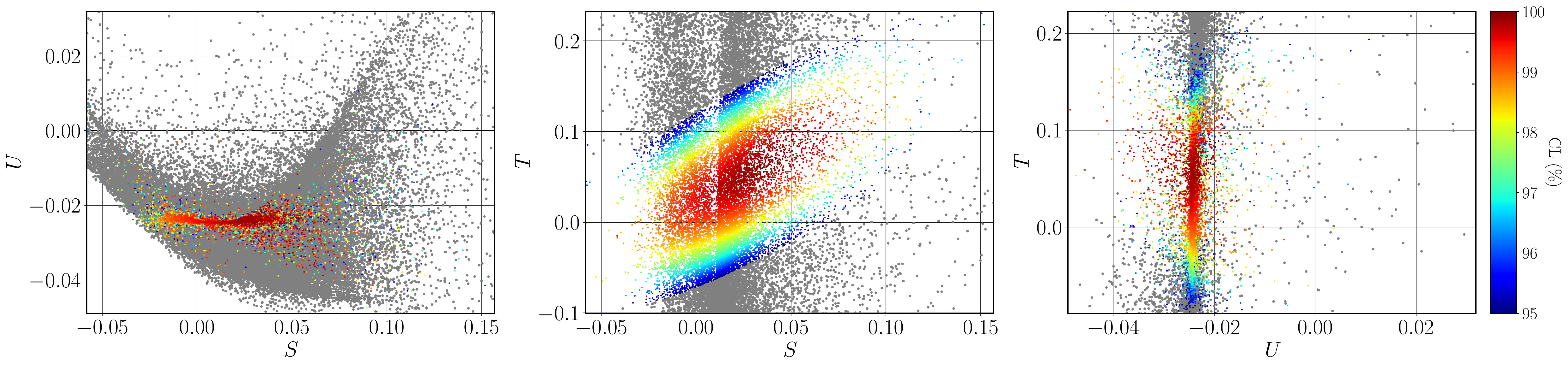}
	\includegraphics[width=0.7\textwidth]{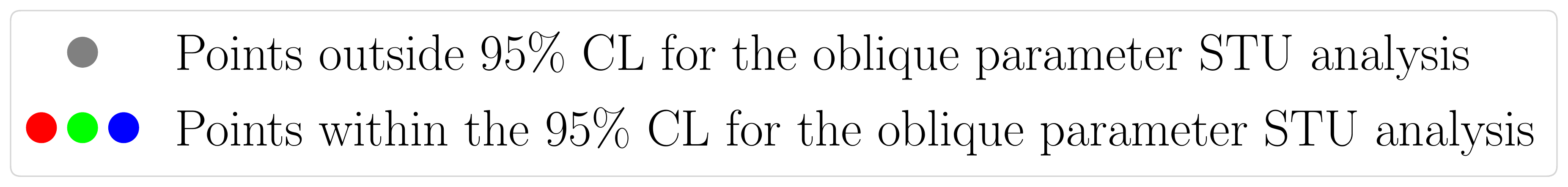}
	\caption{STU electroweak precision observables for all sampled points. Only coloured points pass the STU analysis with a confidence level (CL) of, at least, 95\%.
	}
	\label{fig:STU}
\end{figure}

We show in \cref{fig:STU} the EW precision observables in the $S$-$U$, $S$-$T$ and $U$-$T$ projections. Current precision measurements \cite{Zyla:2020zbs} provide the allowed regions,
\begin{equation}
S = -0.01 \pm 0.10\,, \qquad T = 0.03 \pm 0.12\,, \qquad U = 0.02 \pm 0.11 \ ,
\label{eq:oblique}
\end{equation}
where $S$-$T$ are $92\%$ correlated, while $S$-$U$ and $T$-$U$ are $-80\%$ and $-93\%$ anti-correlated, respectively.
We compare our results with the EW fit in \cref{eq:oblique} and require consistency with the best fit point
%$\Delta \mathcal{O}_{j}^{(0)}$
at $95\%$ C.L. taking into account the correlation between the oblique parameters.

We see that even with our generation of parameters satisfying exact alignment compliance with electroweak precision constraints is not guaranteed, and plenty of points are rejected. However, as expected
in a model with multiple doublets, it is not difficult to find regions of parameter space for which all
constraints on the oblique parameters are satisfied.

\begin{figure}[h!]
	\includegraphics[width=\textwidth]{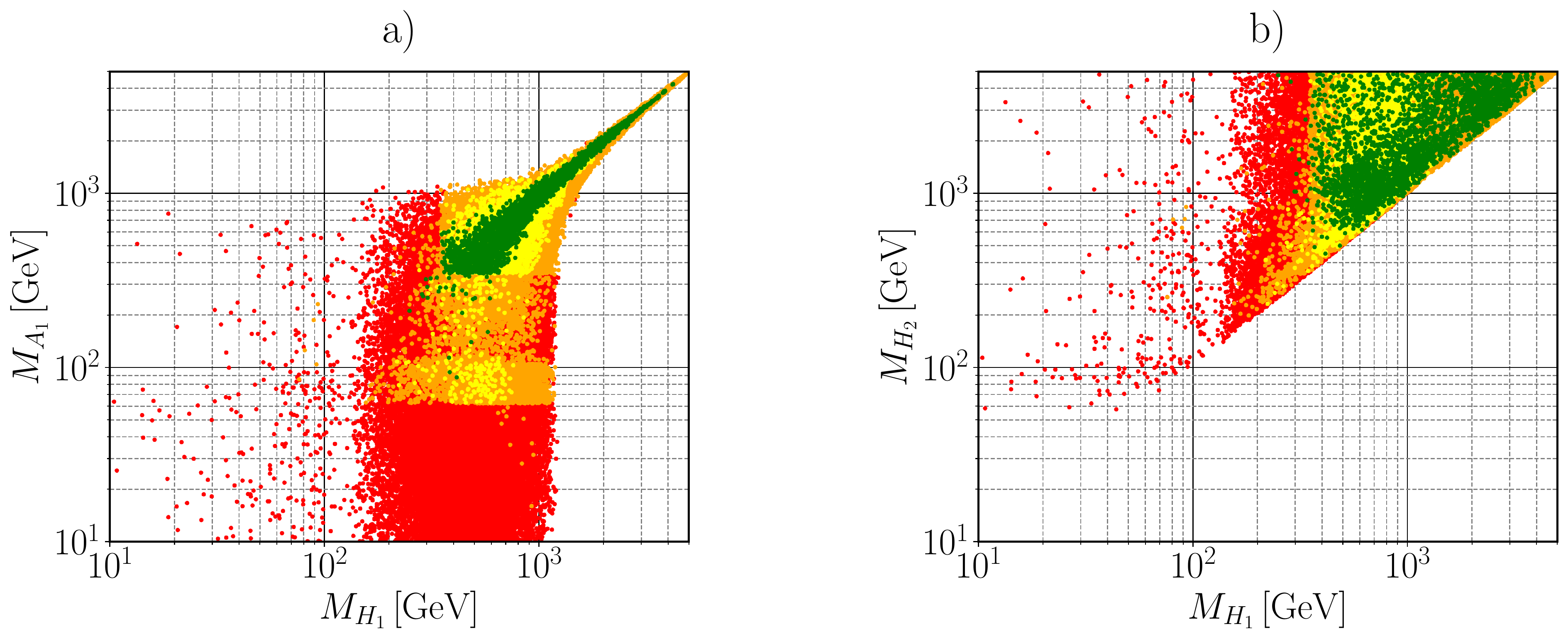}
	\centering
	\includegraphics[width=0.8\textwidth,keepaspectratio,trim= 0 0 0 0]{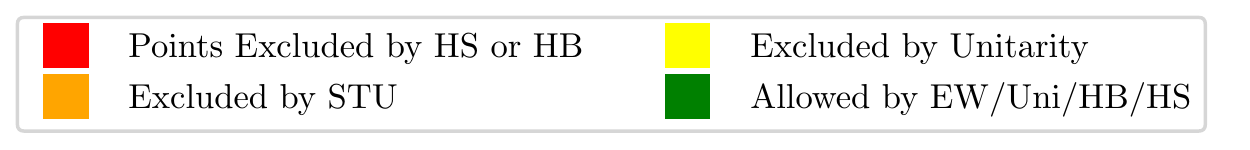}
	\caption{Scatter plots of the allowed parameter space under several constraints imposed by the BGL-like 3HDM. While on the right panel, b), we plot the masses of the two lightest BSM CP-even scalars $H_1$ and $H_2$, the left one, a), showcases the relation between $H_1$ and its heavy CP-even counterpart $H_2$.}
	\label{fig:scamass}
\end{figure}

In \cref{fig:scamass} we show the effect of non-flavour constraints on the allowed parameter space. Here, the impact of restrictions from the LHC experiments, both in terms of measurement of the Higgs bosons'
properties or in the searches for extra scalars, incorporated in the HS/HB framework have been analysed.
Unitarity bounds on the model's quartic couplings are also imposed, as well as precision electroweak constraints via the S, T and U parameters, each leading to a considerable reduction of the allowed parameter space. We see a close correlation between $m_{A_1}$ and $m_{H_1}$ for large values, stemming mostly from unitarity and electroweak precision constraints. Note, the same tendency of near-degeneracy is observed in the mass spectrum of the 2HDM. Furthermore our scan generates very low masses for the scalars, which are excluded by various direct collider searches and implemented in HS/HB. It is important to mention that the size of the input soft masses, together with that of the quartic couplings in \cref{tab:sample} set, approximately, the scale of the physical BSM scalars.

\begin{figure}[h!]
\includegraphics[width=\textwidth,angle=0]{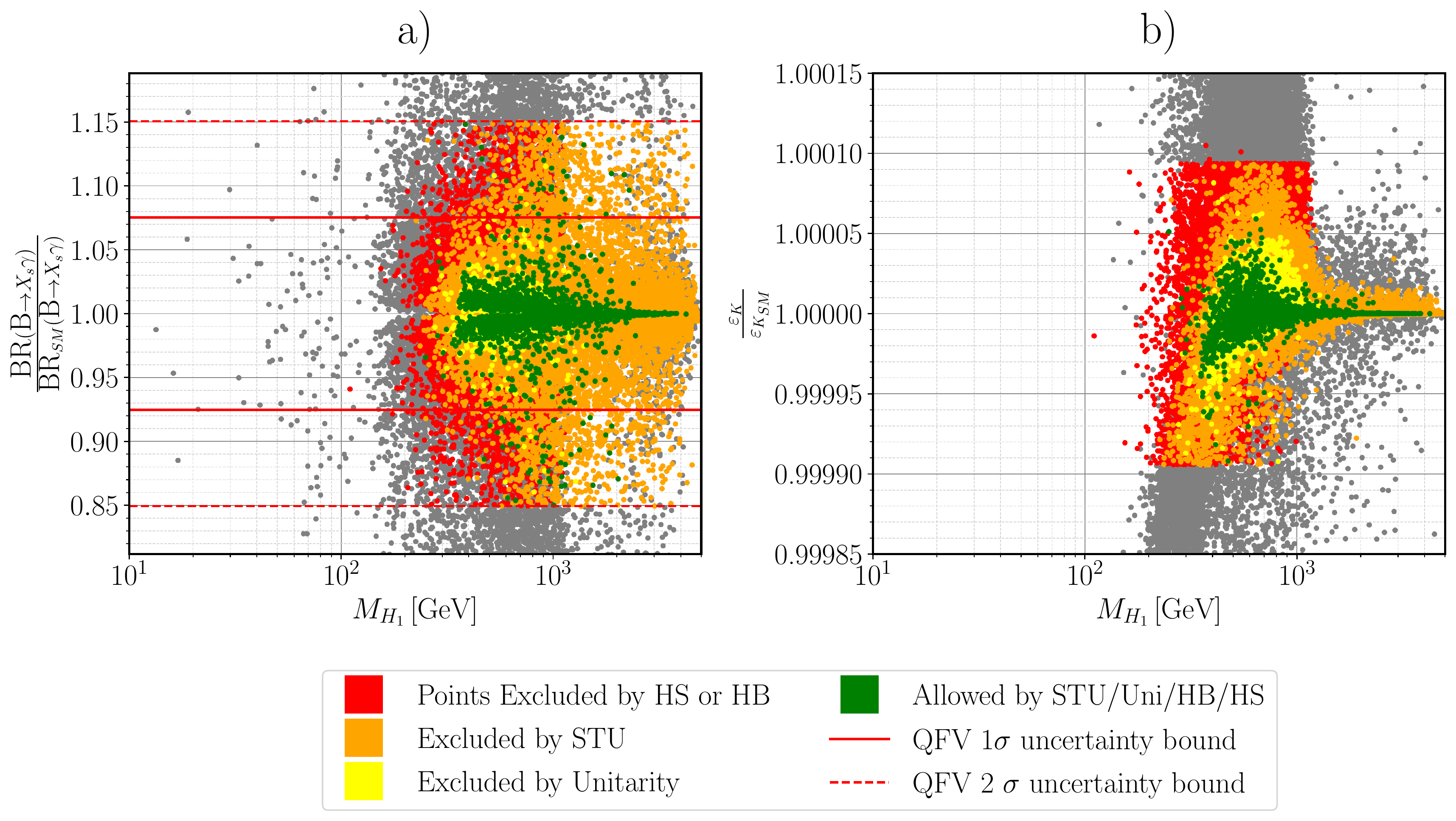}
\caption{Scatter plots of parameter space allowed by several constraints imposed on the BGL-like 3HDM. In the left panel, (a), we show the results for $b\rightarrow s \gamma$, namely,
the ratio of 3HDM-to-SM branching fractions for $B\to X_s \gamma$ reaction while in the right panel, (b),
we plot an analogous ratio for $\epsilon_K$, both in terms of the $H_1$ mass. The colour code is as in \cref{fig:scamass} and grey points are excluded, at 2$\sigma$ level, by at least one QFV observable.
}
\label{fig:flav}
\end{figure}

In \cref{fig:flav} we show how some of the QFV observables might further constrain the model's parameter space that survives the Higgs physics, unitarity and electroweak precision constraints. For instance, the dependency of the ratio of the $b\rightarrow s \gamma$ width computed in the BGL-like 3HDM to the expected SM value as a function of $m_{H_1}$ is shown in \cref{fig:flav}(a). Here, we observe a dispersion around the SM value such that some of the points deviate by more than 2$\sigma$. The $1\sigma$ band is defined in the first line of \cref{tab:err}. In analogy to many known versions of the 2HDM,
the $b\rightarrow s \gamma$ constraint is a very important one, excluding a number of parameter points which otherwise could be perfectly acceptable. Not all flavour variables yield strong constraints, though -- in Fig.~\ref{fig:flav}(b) we show the values obtained within our parameter scan for the Kaon system CP-violating $\epsilon_K$ phase. One notices a rather minuscule variation around the SM value after all other QFV observables have been constrained to lie within a 2$\sigma$ interval of their respective SM-expected values. This is clearly an indication that there are no substantial FCNC contributions to this observable in the considered BGL-like 3HDM.

Note that in \cref{fig:scamass,fig:flav} we have shown only the regions of the parameter space where all constraints from boundedness from below, unitarity, electroweak precision variables and direct searches are obeyed. Having ascertained the relevance of the constraints imposed on the scalar sector, we wish to analyse in detail the impact of the QFV observables, leaving temporarily aside the remaining phenomenological constraints. Since our model has tree-level FCNCs, the inverted procedure described in \cref{sec:Yukawa} does not immediately guarantee a good fit to QFV observables such as the mass differences of the neutral $K$, $B_d$ and $B_s$ mesons, the already mentioned $\epsilon_K$ CP phase, branching ratios such as $B_s\rightarrow \mu^+\mu^-$ etc. Since the SM already does a good job at describing the quark and lepton sector behaviour, we need to verify that NP contributions to those quantities do not ruin the existing agreement between theory and experiment.
\begin{figure}[h!]
\begin{center}
\includegraphics[height=6cm,angle=0]{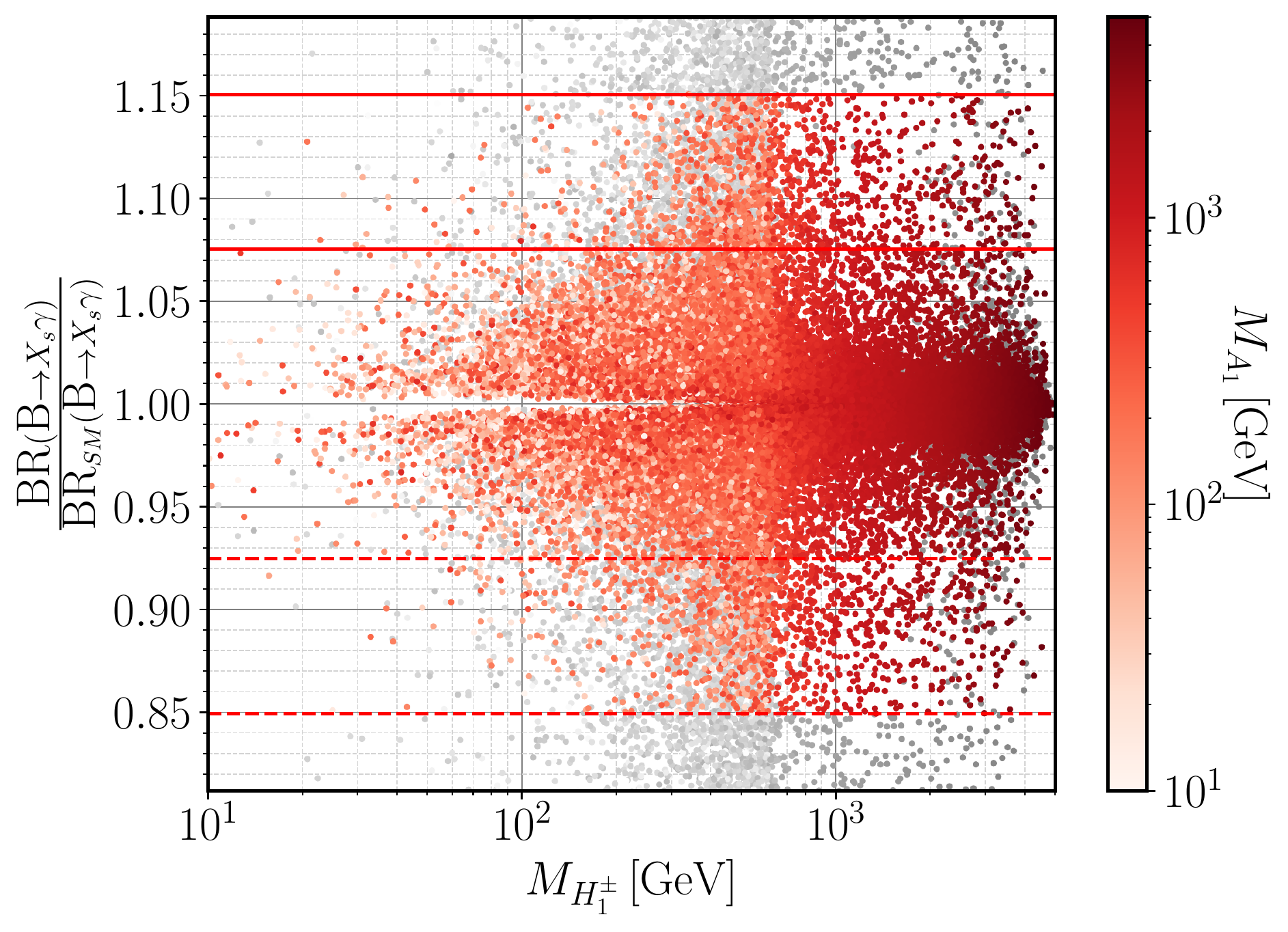}
\includegraphics[height=1.7cm,keepaspectratio]{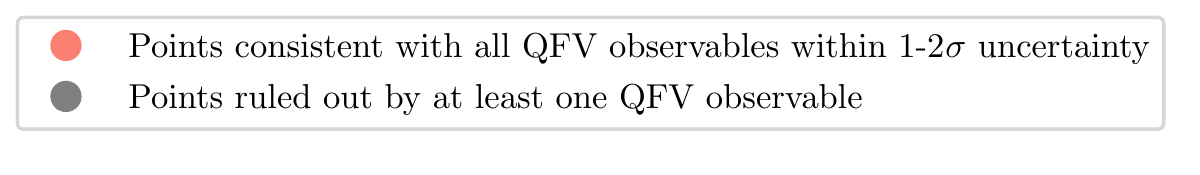}
\end{center}
\caption{The ratio of 3HDM-to-SM branching fractions for $B\to X_s \gamma$ decay in the considered BGL-like 3HDM as a function of the $H^\pm_1$ mass. Red-shaded points are those for which all QFV constraints are satisfied at most at 2$\sigma$ (many even at 1$\sigma$) level, grey points are those for which at least one QFV observable is in disagreement with the current measurements by more than 2$\sigma$. The horizontal lines account for the 1 and 2$\sigma$ uncertainties as in the first line of \cref{tab:err}. The ``temperature" gradient of colour shows the lightest pseudoscalar mass.
 }
\label{fig:charga}
\end{figure}

As we have already mentioned, our model, being BGL-like, ought to allow for an easy fit in the flavour sector, an assumption we now put to the test. In Fig.~\ref{fig:charga} we see how an agreement with $b\rightarrow s\gamma$ constraints is not automatic -- in fact, we observe a number of (grey) points with larger than 2$\sigma$ deviation from the SM prediction, and quite a few disagreeing with the SM $b\rightarrow s\gamma$ values between 1 and 2$\sigma$. These larger than 1$\sigma$ deviations are seen to occur mostly (though not exclusively) for lower masses of the lightest charged scalar, $m_{H_1}^\pm \leq 700$ GeV.
This is similar to what occurs for the Type II 2HDM~\cite{Arbey:2017gmh}.
Since in our model the down-type quark masses do not arise from a single $\Gamma$ matrix, it is in fact natural that we find regions of parameter space for which observables such as $b\rightarrow s\gamma$ behave in a similar manner to a Type II 2HDM. However, we see that our 3HDM can fit this observable for much lower charged Higgs masses than 700 GeV, as occurs, for instance in a Type I 2HDM -- again to be expected, certain regions of our parameter space should mimic well the Type-I behaviour.
A similar phenomenon was observed for a 2HDM with tree-level FCNCs, see~\cite{Ferreira:2019aps}.

We further observe that the values of the $B\to X_s \gamma$ width in our model approach the corresponding SM value for very large values of the lightest charged Higgs boson mass\footnote{As we saw in \cref{fig:scamass}(b), theoretical and experimental constraints imposed upon the model force the extra scalars to have small mass splittings for large values of their mass. A value of $m_{H_1^\pm}$ above 1 TeV thus corresponds to all other scalar particles having masses of the same order.}. This is not surprising since NP contributions to this observable depend on the inverse of the square of the extra scalars' masses and are thus expected to approach zero as those masses tend to infinity.
\begin{figure}[h!]
\begin{center}
\includegraphics[height=6cm,angle=0]{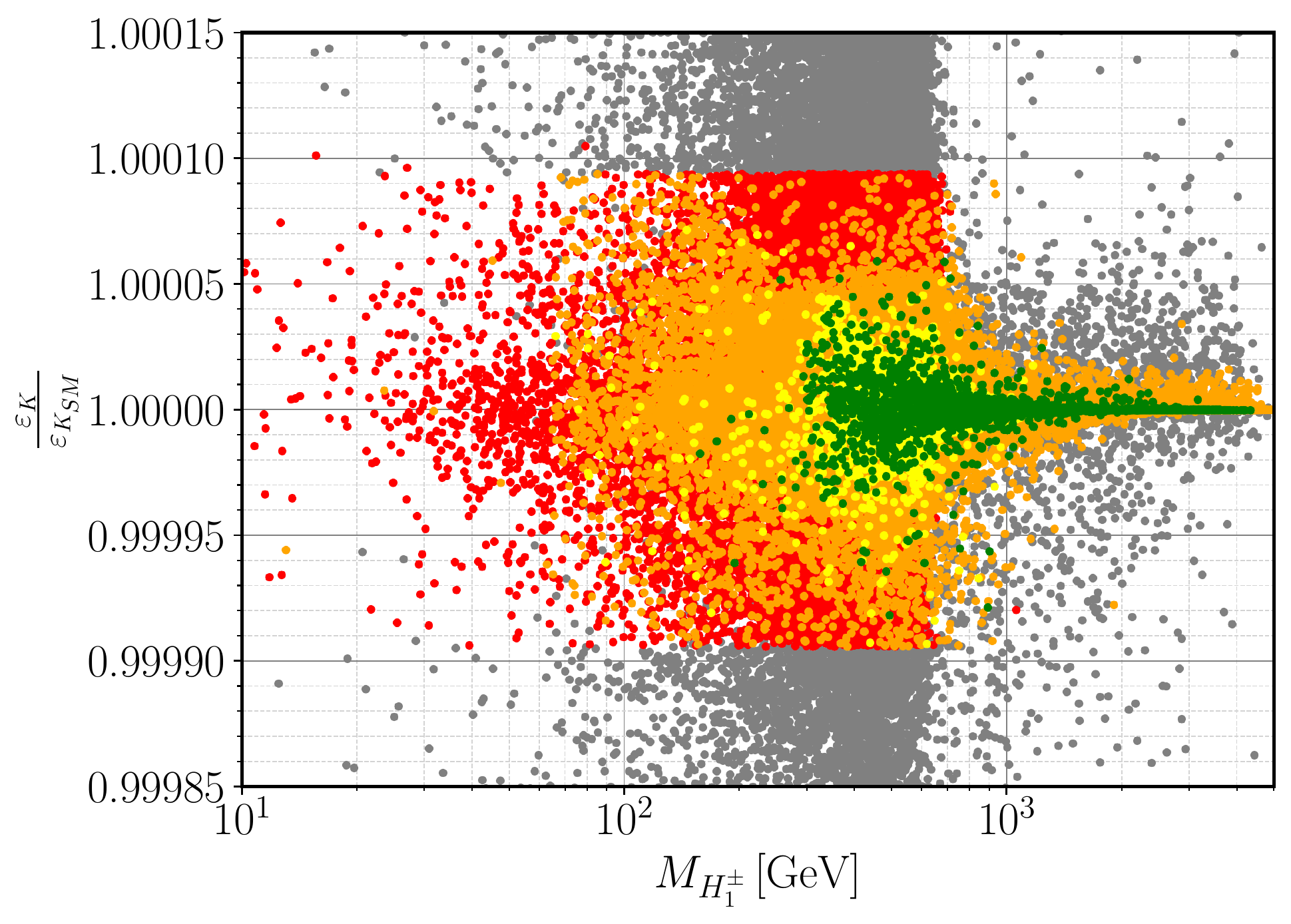}
\end{center}
\caption{$\epsilon_K$ as a function of the $H_1^\pm$ mass for the parameter space allowed under several constraints imposed on the BGL-like 3HDM. The colour code is the same as in \cref{fig:flav}.
}
\label{fig:chargb}
\end{figure}
In Fig.~\ref{fig:chargb} on the other hand, considering again the full set of phenomenological constraints, we observe how the inverted procedure we are using ~to constrain the Yukawa sector yields an excellent agreement with other QFV observables -- there we plot the values of $\epsilon_K$ as a function of the lightest charged Higgs boson mass, and see how close it gets to the SM value for all the generated points. We see that this observable attains, in this model, values extremely close to the SM prediction, with deviations of the order of $\sim$ 0.01\%. To put these results in context, the current experimental uncertainty on $\epsilon_K$ stands at less than 0.5\% of its central value. The minimal value of the charged Higgs boson mass that still reproduces the experimental value of $\epsilon_K$ and satisfies all constraints is found to be $\sim 150$ GeV.

\begin{figure}[h!]
\begin{center}
\includegraphics[height=6cm,angle=0]{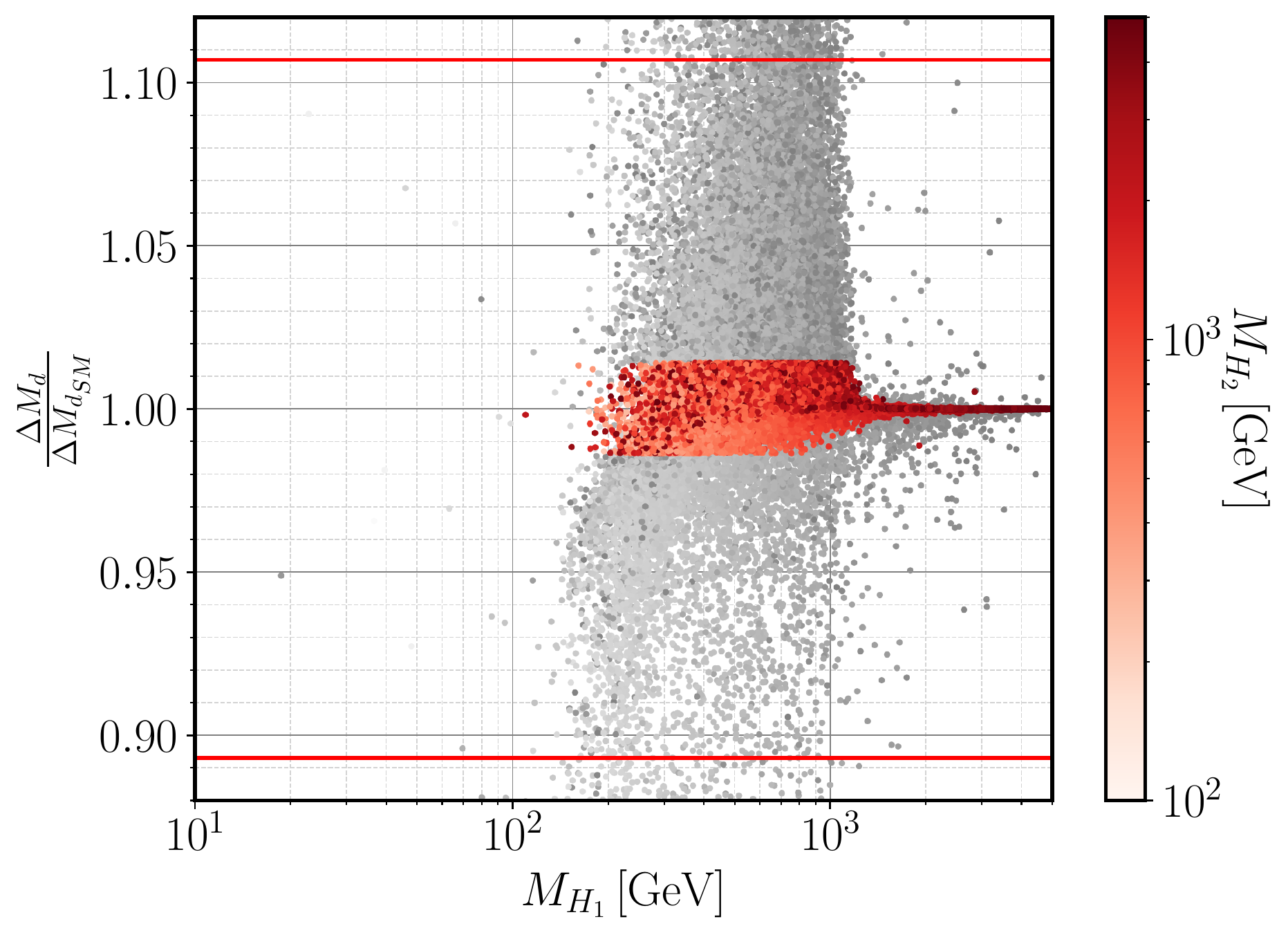}
\end{center}
\caption{$B_d$ mass difference as a function of
heavier CP-even Higgs boson masses. The colour code is the same as in \cref{fig:charga}.
}
\label{fig:neutb}
\end{figure}

For completeness, let us also consider the $B$-meson mass differences. These are the observables which in the SM are generated by one-loop box diagrams but also receive tree-level contributions in theories with scalar mediated FCNC interactions in the down-quarks sector.
\begin{figure}[h!]
\begin{center}
\includegraphics[height=6cm,angle=0]{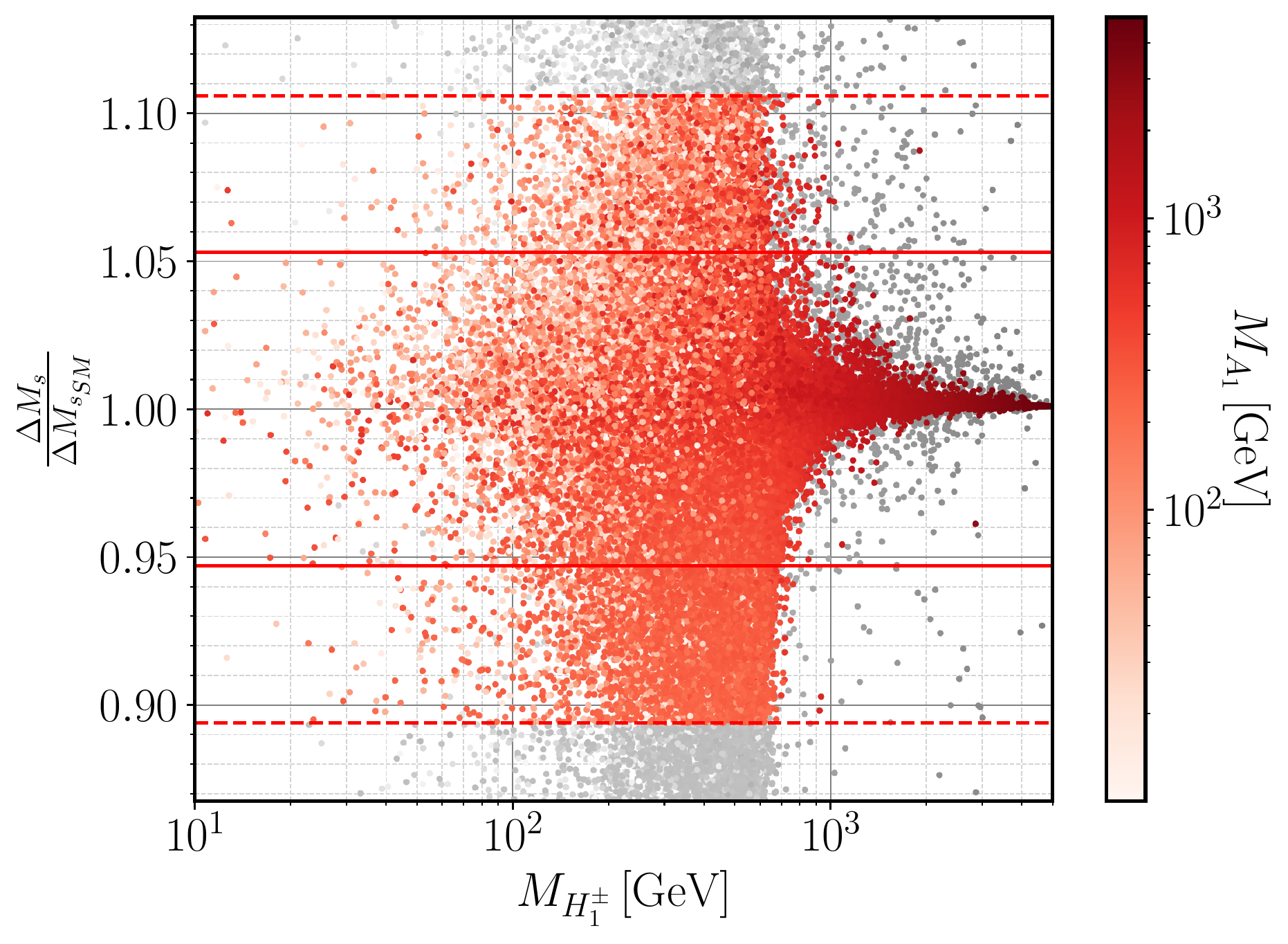}
\end{center}
\caption{$B_s$ mass difference as a function of the the CP-even $H_1$ and pseudoscalar $A_1$ masses.
The colour code is as in Fig.~\ref{fig:charga}.
}
\label{fig:neuta}
\end{figure}
Again, and as expected, we see in \cref{fig:neuta,fig:neutb} that the values obtained in the BGL-like 3HDM for $\Delta M_{B_s}$ and $\Delta M_{B_d}$ approach their SM values for large enough masses of the extra scalars. We also see that our scanning procedure produces values of $\Delta M_{B_d}$ extremely close to that of the SM (even for lower masses), with a larger dispersion found in $\Delta M_{B_s}$, still within a 2$\sigma$ variation. This is clearly due to the fact that we chose a specific structure for the Yukawa matrices in \cref{e:ndf} in order to
single out the third generation. Furthermore, for a BGL-like model, the FCNC interactions are expected to be suppressed by the CKM matrix elements, which, for the B-meson oscillation observables under consideration, explains how contributions to $\Delta M_{B_d}$, which involve a ``jump" across two generations, are more suppressed than those contributions to $\Delta M_{B_s}$, for which scalars only ``jump" one generation in their QFV interactions.

While we do not show all the numerical results explicitly, we have analysed a wealth of other flavour physics observables, encountering 1$\sigma$ agreement with current experimental bounds for all of them. These included the remaining QFV observables such as neutral Kaon mass
differences, neutral $B$ mesons decays to muon and electron pairs and other leptonic sector measurements, $Z\rightarrow b\bar{b}$ observables etc.

\subsection{A degree of fine-tuning}

The remarkable agreement found in the previous section for QFV observables, even when including the FCNC contributions from the extra scalars present in our model, needs to be analysed in depth. Such an agreement can arise in several ways. For instance, if all extra scalar boson masses are very high, then the NP contributions take very small values and an agreement with the SM value is easily found. Another possibility is that the FCNC Yukawa interactions are {\em naturally} small, something which occurs in the 2HDM BGL model. Their strong suppression follows from the small off-diagonal CKM matrix elements as a consequence of the symmetry of the model, which we claim to also happen in the current 3HDM version. And finally, there is also the possibility of a fine tuning having occurred in the scanning procedure, ``unnaturally" causing cancellations between different NP contributions. Let us now demonstrate that this last possibility does not occur in our model.

In order to see an example of such possible fine tuning, consider the Kaon system and observables such as the $K^0-\bar{K}^0$ mass difference, or the CP-violating phase $\epsilon_K$. The matrix element describing the transition $\bar K^0 \to K^0$, $M_{21}$, receives contributions from the SM, via box diagrams, and from NP, through tree-level FCNCs in the scalar sector:
$M_{21} = M_{21}^\mathrm{SM} + M_{21}^\mathrm{NP}$. The NP terms arise in our model from tree-level Feynman diagrams and thus can in principle overwhelm the SM result. These diagrams represent the tree-level exchanges of CP-even and CP-odd scalars with FCNC interactions which, using the vacuum-insertion approximation
(see Refs.~\cite{Branco:1999fs,Ferreira:2011xc,Ferreira:2019aps}), are found to be
\ba
M_{21}^\mathrm{K,NP}
&=& \frac{f_K^2 m_K}{96 v^2} \left\{
\frac{10 m_K^2}{\left( m_s + m_d \right)^2}
\left( \frac{F^{dA_1}_{ds}}{m_{A_1}^2} + \frac{F^{dA_2}_{ds}}{m_{A_2}^2}
- \frac{F^{dh}_{ds}}{m_h^2}
- \frac{F^{dH_1}_{ds}}{m_{H_1}^2} - \frac{F^{dH_2}_{ds}}{m_{H_2}^2}
\right)
\right. \nonumber \\
& & \left.
+ 4 \left[ 1 + \frac{6 m_K^2}{\left( m_s + m_d \right)^2} \right]
\left( \frac{\bar{F}^{d A_1}_{ds}}{m_{A_1}^2} + \frac{\bar{F}^{d A_2}_{ds}}{m_{A_2}^2}
+ \frac{\bar{F}^{d h}_{ds}}{m_h^2}
+ \frac{\bar{F}^{d H_1}_{ds}}{m_{H_1}^2} + \frac{\bar{F}^{d H_2}_{ds}}{m_{H_2}^2}
\right)
\right\} \,,
\label{eq:M21K}
\ea
where $f_K $ and $m_K$ are the $K$-meson decay constant and mass, respectively,
and the following combinations of FCNC couplings were defined, for each scalar/pseudoscalar
$X = \{h, H_1, H_2 , A_1, A_2\}$, as follows
\ba
F^{dX}_{ab} &=& \left(N^*_{dX}\right)^2_{ab}\,+\,\left(N_{dX}\right)^2_{ba}\,, \nonumber \\
\bar{F}^{dX}_{ab} &=& \left(N^*_{dX}\right)_{ab}\left(N_{dX}\right)_{ba}\,.
\ea
Here, the matrices $N_{dX}$ are the Yukawa matrices for down-type quark interactions with each scalar $X$. We observe in Eq.~\eqref{eq:M21K} that CP-even contributions tend to cancel CP-odd ones. This opens up the possibility to fit the Kaon observables if the CP-even and CP-odd
terms, though potentially large by themselves, cancel up to the n$^{th}$ decimal place. Therefore, such a cancellation would produce a result that, while in nominal agreement with experiments, is ``unnatural", and would be potentially challenged by higher order corrections. Similar fine-tunings should be investigated in other observables, such as the $B$-meson mass differences, or semileptonic quantities such as the branching ratio of $B_s\rightarrow \mu^+\mu^-$ etc.

\begin{figure}[h!]
\begin{tabular}{cc}
\includegraphics[height=6cm,angle=0]{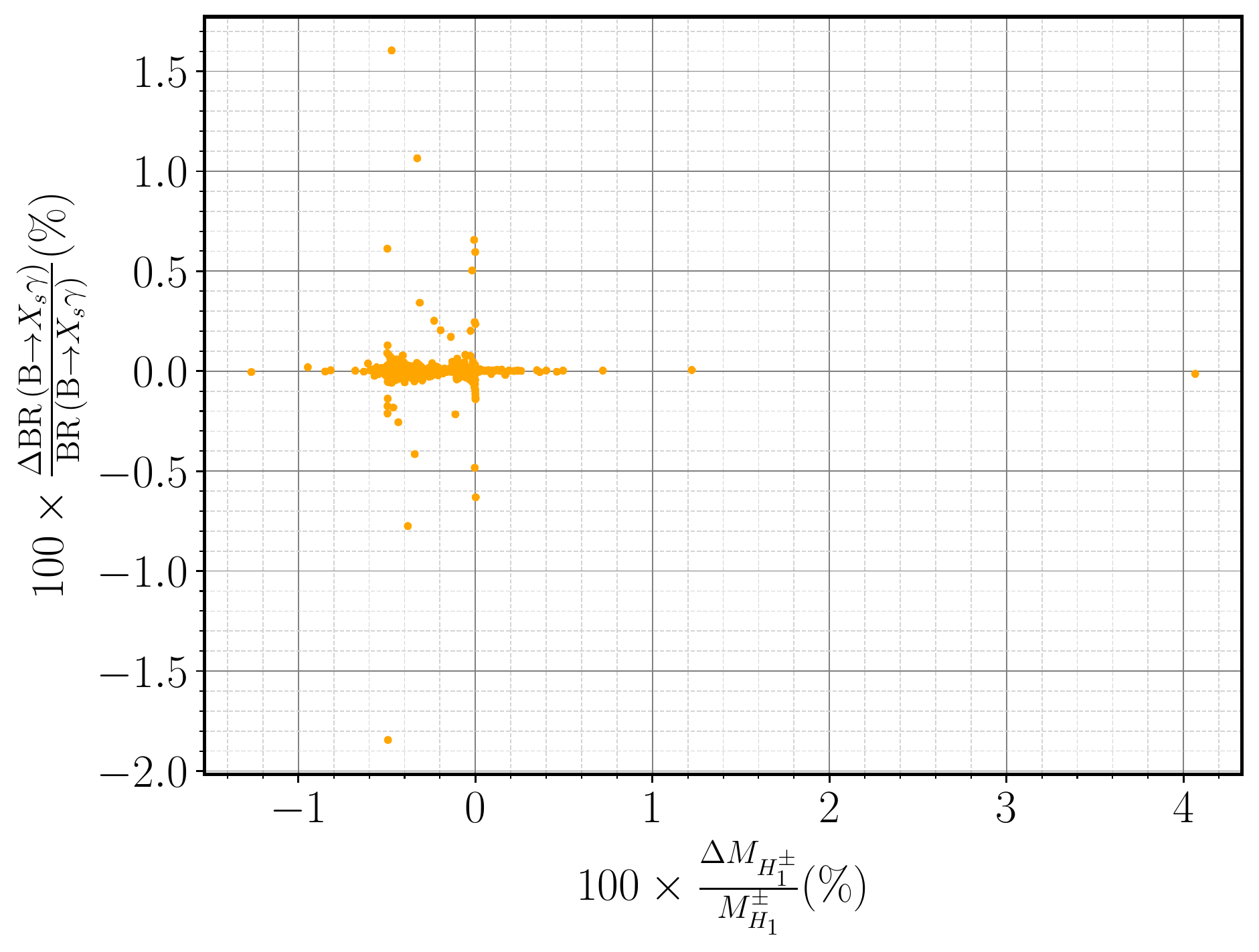}&
\includegraphics[height=6cm,angle=0]{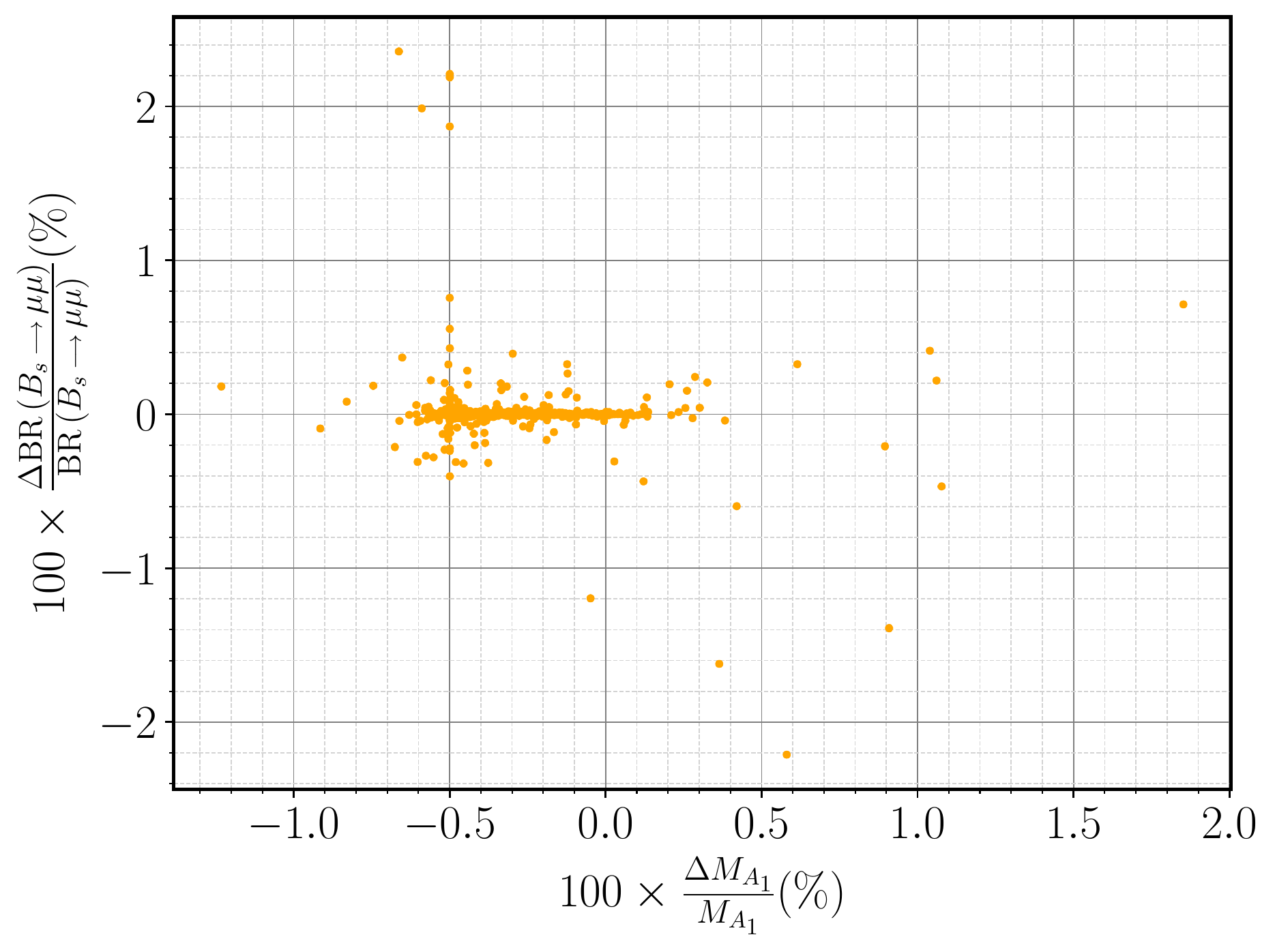}\\
 (a) & (b)
\end{tabular}
\caption{Percentage of variation of the branching fractions (a) for $B\to X_s \gamma$ decay and (b) for $B_s\rightarrow \mu^+\mu^-$ decay as functions of the percentage of variation in scalar boson masses. In the set of input parameters for which all experimental and theoretical constraints were satisfied, each of the soft breaking parameters were smeared by less than 1\% around their initial values, keeping the Yukawa couplings and mixing angles fixed.}
\label{fig:fint}
\end{figure}

In order to investigate whether our results were fine-tuned or not,
we have undertaken the following procedure:
\begin{itemize}
\item Chose a combination of parameters for which all theoretical and experimental constraints are satisfied. Here, we were focused on the parameter space points that give rise to the predicted values for QFV observables that are at most within 2$\sigma$ of their SM counterparts.
\item Fixed all the model parameters except for the soft breaking mass terms, which were then smeared with a random variation of no more than 1\% about their initial values. The scalar mass spectrum was then re-calculated and we chose those situations for which there were variations of less than 5\% on all scalar boson masses.
\item With the new values of the scalar boson masses (and all remaining mixing angles and Yukawa couplings being fixed to the values prior to the smearing procedure described above) we recalculate the QFV observables and compare them with their initial values.
\end{itemize}
If there is a fine tuning in the calculation of $\epsilon_K$, for instance, a small variation in the masses, such as $m_{A_1}$ or $m_{H_1}$, should lead to a much larger variation of the observable. We see the results of this procedure in \cref{fig:fint}.
Here we show variations observed for the branching ratios of $B\to X_s \gamma$ and $B_s\rightarrow \mu^+\mu^-$ decays. Similar variations were found for other observables, such as the meson mass differences $\Delta M_d$ and $\Delta M_s$ or the branching ratio for $B_s\rightarrow \mu^+\mu^-$ decay. Other observables had by far smaller variations -- for
comparison, the absolute value of variations on $\epsilon_K$, for instance, were always found to be less than 0.01\%.

Thus, we observe that, while keeping all mixing angles and Yukawa couplings the same, smearing the physical scalar masses by up to 4\% around their starting values for phenomenologicaly viable points, induces variations of less than 2.5 \% on the values of typical QFV observables. We therefore conclude that in our BGL-like 3HDM the best parameter space points that pass all the considered constraints are not the result of an accidental fine-tuning, rather {\em the FCNC interactions of the model have been rendered naturally small by the symmetries of the model}, just as it occurs in the standard 2HDM BGL.

\subsection{LHC predictions}
\label{sec:lhc}

As we saw in the previous sections, the BGL-like 3HDM under discussion can fit, despite the presence of tree-level FCNC interactions in the down sector, all current experimental constraints without any fine tuning. Furthermore, it can do so even with extra scalar masses below $\sim$ 500 GeV raising a tantalizing question: could such scalars have already been observed at the LHC, or can the current LHC data be used to exclude portions of the model's parameter space? Besides, what can one expect {\em vis-\`a-vis} future LHC sensitivity to potentially discover the extra scalars predicted in our model?

To begin with, we must recall that we have chosen to work in the exact alignment limit in this model. Thus, the CP-even scalars $H_1$ and $H_2$ have couplings to $Z$ (and $W$) boson pairs exactly equal to zero. As such, experimental searches for extra scalar resonances decaying into $Z$ or $W$ boson pairs~\cite{Sirunyan:2018qlb,Sirunyan:2019pqw,Aad:2020ddw} are automatically satisfied in our parameter scan. Deviations from the alignment limit to be considered in a more general analysis might somewhat change that state of affairs. Given how SM-like the 125 GeV scalar appears to be in the LHC measurements, the couplings of $H_1$ and $H_2$ to electroweak gauge bosons should always be heavily suppressed,
so we do not perform such more generic off-alignment analysis in the current work. However, the sum rule for scalar-gauge boson couplings -- which in the exact alignment limit makes $H_1$ and $H_2$ gauge-phobic -- does not apply to Yukawa interactions. For instance, in the 2HDM or in SUSY models, interactions of the pseudoscalar $A$ to fermion pairs may
be enhanced (or suppressed) by a factor of $\tan\beta$. A promising avenue in searches for additional scalars is therefore the di-tau channel, where the current and future LHC sensitivities may well reveal their presence.

\begin{figure}[h!]
\begin{center}
\includegraphics[height=6cm,angle=0]{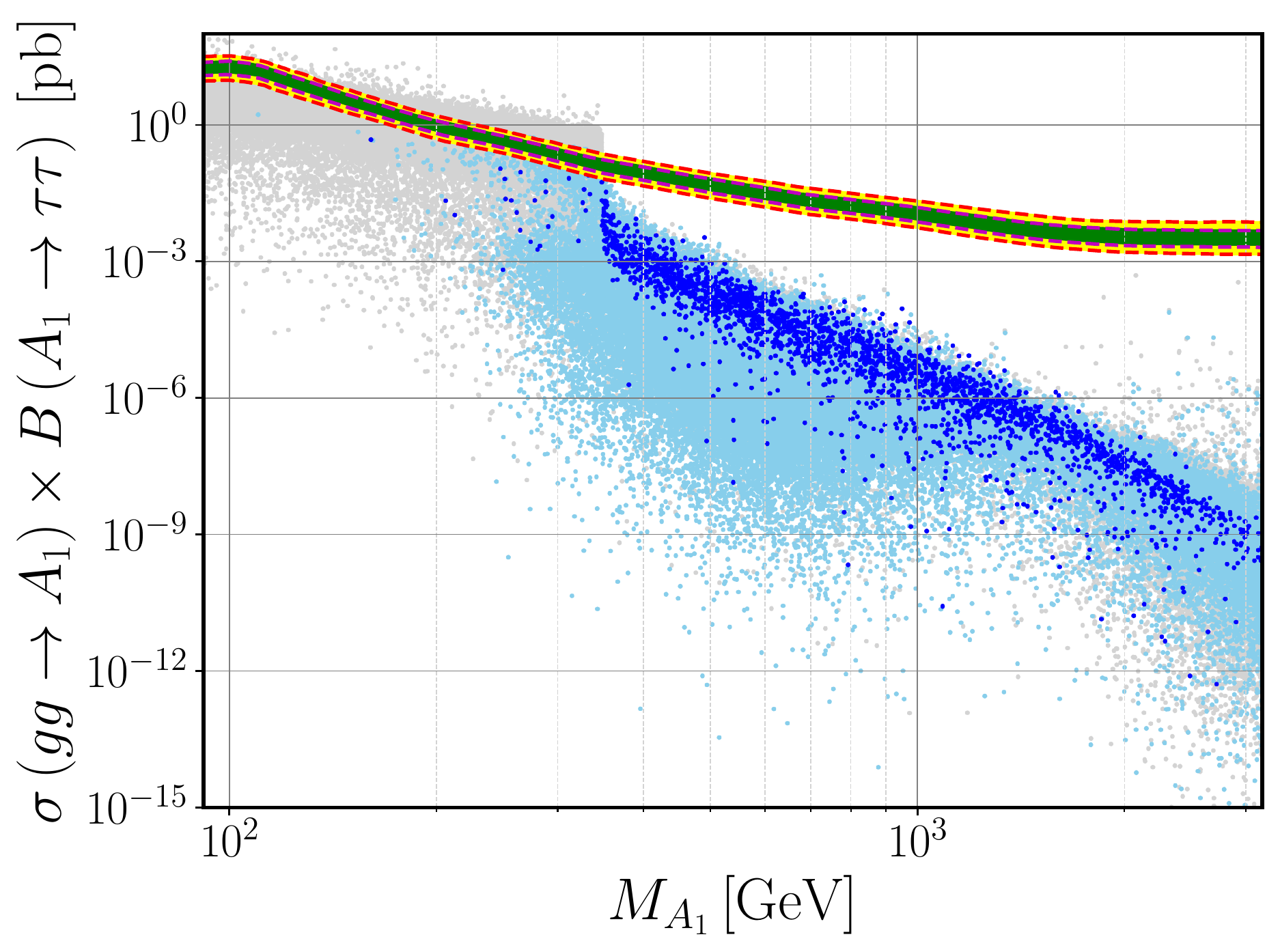}
\includegraphics[height=3cm,angle=0, trim= 0 -1cm 0 0]{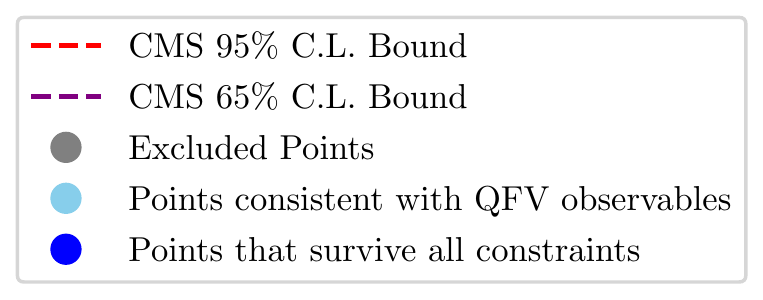}
\end{center}
\caption{The pseudoscalar $A_1$ production cross section in gluon-gluon fusion at the LHC center of mass energy of 13 TeV, times its decay branching ratio to $\tau\bar{\tau}$ as a function of $m_{A_1}$. Grey points represent the sets of scenarios excluded by not obeying at least one QFV observable at $2 \sigma$. Blue points, both dark and light shades, correspond to an agreement with all QFV observables at least at 2$\sigma$ level. Those points that are further allowed by all imposed constraints are represented by the dark blue points. The 1 and 2$\sigma$ observation limits available from the CMS Collaboration for searches in this channel are taken from~\cite{Sirunyan:2018zut}.
}
\label{fig:Atau}
\end{figure}

In \cref{fig:Atau} we show the cross section for the production of the lightest pseudoscalar $A_1$ in a gluon-gluon fusion process multiplied by its branching ratio into tau pairs, for a center-of-mass energy of 13 TeV. The CMS $65\%$ and $95\%$ exclusion bounds were extracted from \cite{Sirunyan:2018zut}.
The pseudoscalar production cross section via gluon-gluon fusion was obtained using \texttt{MadGraph5\_aMC@NLO 2.6.2} \cite{Alwall:2014hca}. We have used an interface with \texttt{SPheno} where masses, decay widths and branching fractions are calculated and linked to \texttt{MadGraph5} for every single generated point. The relevance of requiring an agreement with QFV observables is emphasized in this plot, where grey points represent scenarios where at least one of such observables was found to deviate from its SM value by more than 2$\sigma$. The blue points denote scenarios for which a full 1$\sigma$ agreement was found for all QFV observables while the darker shades of blue also survive all remaining theoretical and experimental constraints. As was to be expected, the total signal strength for this channel diminishes as the pseudoscalar mass increases -- see in particular the sharp drop-off around $m_{A_1} \sim 375$ GeV, that is twice the top mass. Indeed, for larger pseudoscalar
masses, a new decay channel $A_1 \rightarrow t\bar{t}$ becomes kinematically allowed and tends to reduce the branching ratio for $A_1 \rightarrow \tau\bar{\tau}$. Notice, however, that there is a number of points for lower $m_{A_1}$ which can almost be probed by the current CMS bounds.

Until the end of LHC operation we can expect an increase in accumulated luminosity by at least a factor of 100, which would roughly lower the exclusion lines shown in \cref{fig:Atau} by an order of magnitude. As such, we can expect the searches in this channel to at least exclude parts of the parameter space for $m_{A_1} < 400$ GeV. In fact, we see in \cref{fig:Atau} that the maximum of the signal strength occurs for $m_{A_1} \simeq 350$ GeV, which is unsurprising, given that this value roughly corresponds to twice the top mass. In fact, it is well known that the gluon-gluon fusion cross section has a local maximum for a c.o.m. energy equal to twice the top mass, both for the production of a CP-even or a CP-odd scalar.

\begin{figure}[t]
\begin{tabular}{cc}
\includegraphics[height=6cm,angle=0]{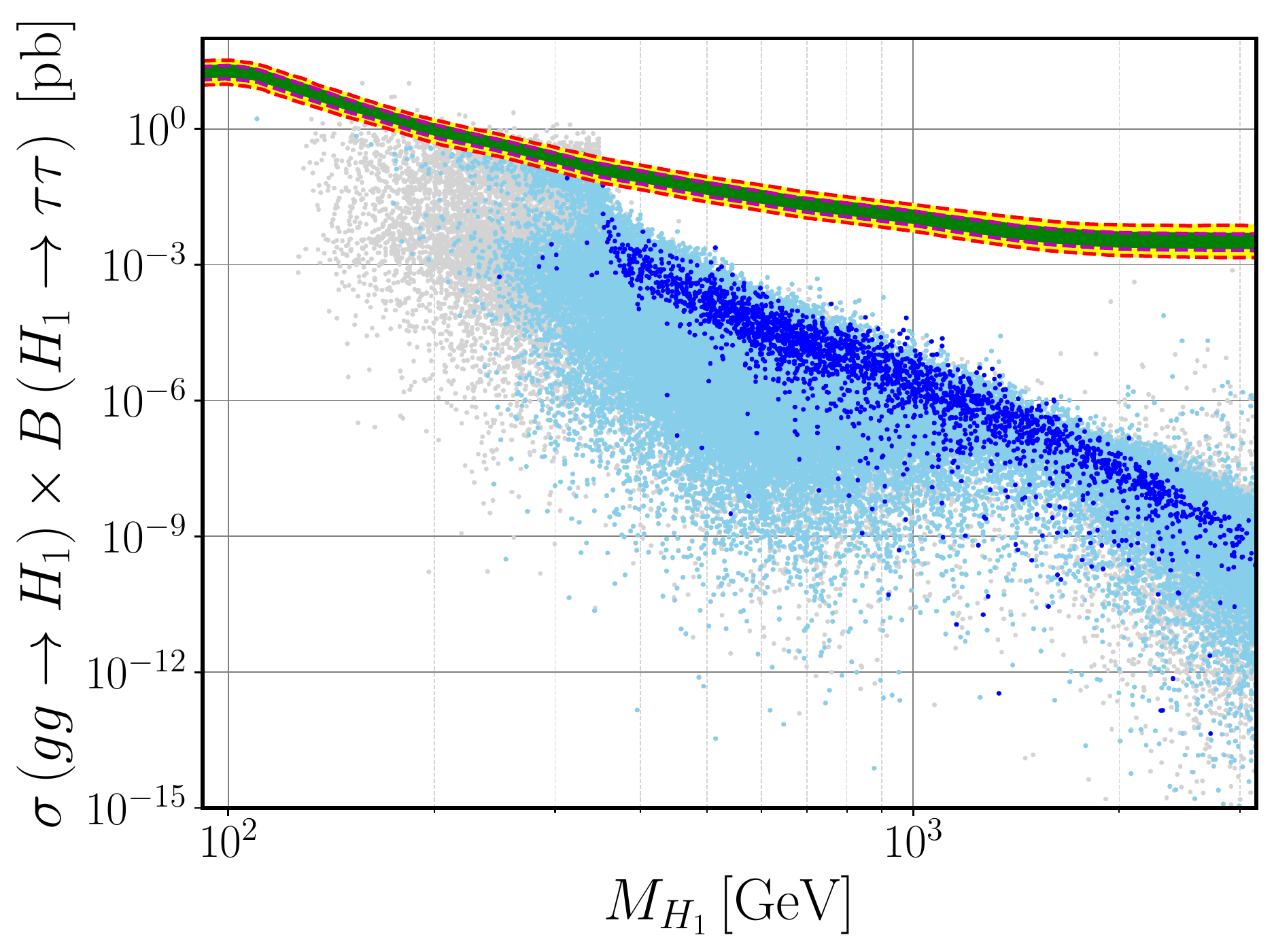}&
\includegraphics[height=6cm,angle=0]{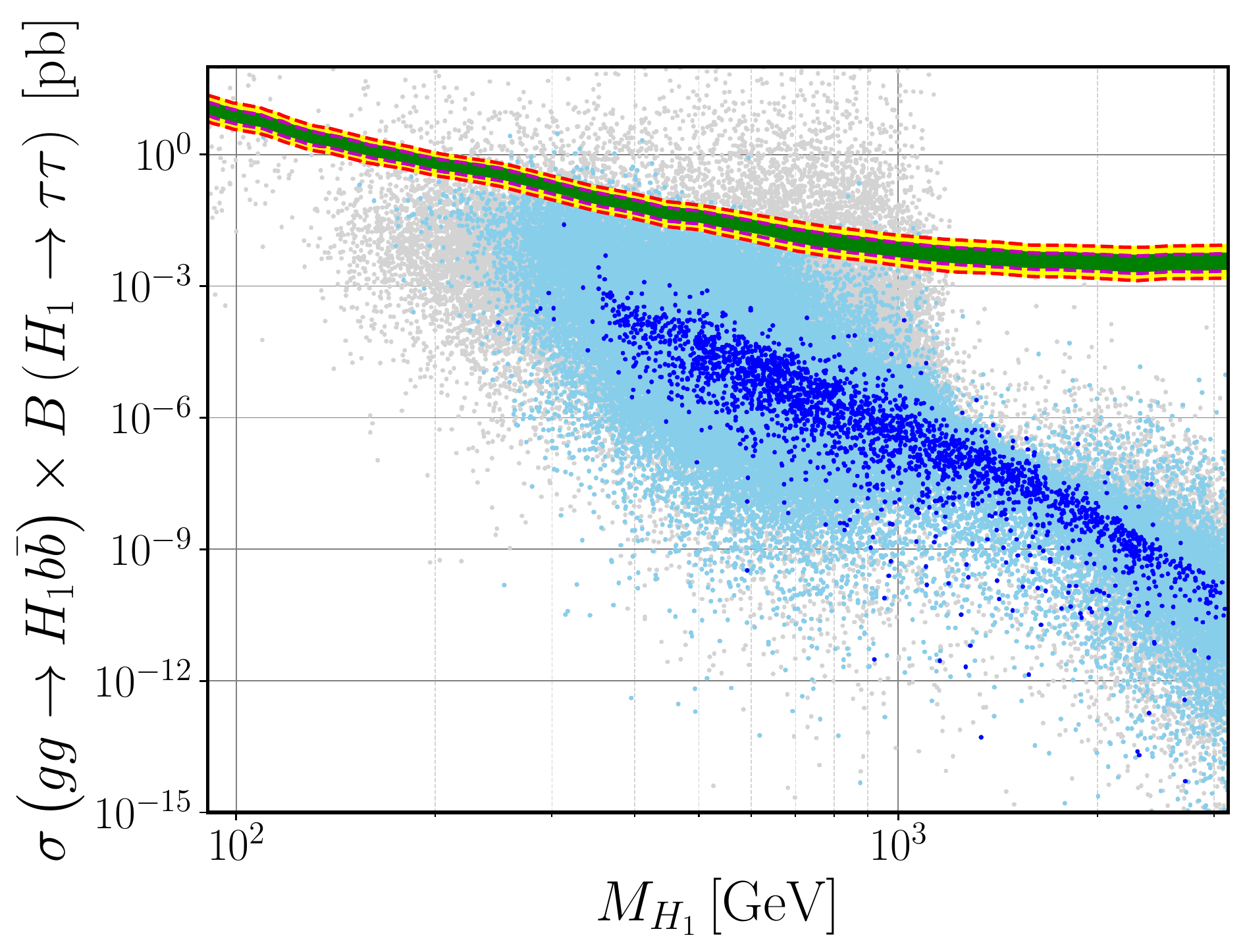}\\
 (a) & (b)
\end{tabular}
\caption{The signal strength for production of a CP-even scalar via the gluon-gluon fusion mechanism (a) times its branching ratio to $\tau\bar{\tau}$, and (b) with associated production of a $b\bar b$-pair
times its branching fraction to $\tau\bar{\tau}$, as a function of the lightest CP-even mass, $m_{H_1}$. The colour code is the same as in \cref{fig:Atau} and the exclusion bounds in both panels were also taken from \cite{Sirunyan:2018zut}.
}
\label{fig:Htau}
\end{figure}
The di-tau channel is also appropriate in searches for a heavier CP-even state, as we see in \cref{fig:Htau}. As before, take notice of the expected sharp drop in the value of the signal rate for masses $m_{H_1} > 2 m_t$. Both in direct production via gluon-gluon fusion into $H_1$, or in its associated production with a bottom quark pair, the obtained signal strength including the branching ratio for $H_1 \rightarrow \tau\bar{\tau}$ is very close to the current CMS sensitivity for the lower mass region. Thus, we see that our BGL-like 3HDM is close to being probed by the current LHC data, and before the end of the next LHC run certain parts of its parameter space can also be tested in direct searches for BSM scalars. We therefore provide five representative benchmark points in \cref{tab:bench} to be searched for in the LHC run-III. These were chosen such that they obey all theoretical and experimental contraints on the scalar, gauge and fermion sectors, and further satisfying the following criteria:
\begin{table}[]
\centering
\begin{tabular}{ccccccc}
\toprule
    & $m_{H_1}$ & $m_{H_2}$ & $m_{A_1}$ & $m_{A_2}$ & $m_{H^\pm_1}$ & $m_{H^\pm_2}$ \\ \midrule
\textbf{BP1} & 249 & 4746 & 212 & 4740 & 101 & 4736 \\ \midrule
\textbf{BP2} & 285 & 4364 & 264 & 4419 & 146 & 4371  \\ \midrule
\textbf{BP3} & 576 & 695 & 161 & 930 & 484 & 663 \\ \midrule
\textbf{BP4} & 437 & 702 & 206 & 881 & 380 & 710 \\  \midrule
\textbf{BP5} & 452 & 659 & 338 & 723 & 420 & 793 \\  \midrule
\textbf{BP6} & 313 & 1564 & 326 & 1457 & 230 & 1469 \\  \midrule
\textbf{BP7} & 353 & 1067 & 582 & 1056 & 312 & 1091 \\ \bottomrule \vspace{1cm}
\end{tabular}
\begin{tabular}{cccccc}
\toprule
\multicolumn{1}{l}{} & $ \frac{\textrm{BR} ( \textrm{B} \rightarrow X_s \gamma)}{\textrm{BR}_{SM} ( \textrm{B} \rightarrow X_s \gamma )}$ & $ \frac{ \textrm{BR} (B_s  \rightarrow  \mu  \mu )}{\textrm{BR}_{SM}(B_s  \rightarrow  \mu  \mu ) }$ & $ \frac{\Delta M_d}{\Delta M_{d_{SM}} }$ & $ \frac{\Delta M_s}{\Delta M_{s_{SM}} }$ & $ \frac{\varepsilon_K}{\varepsilon_{K_{SM}}} $ \\  \midrule
\textbf{BP1} & 0.98 & 1.01 & 1.00 & 0.99 & 1.00 \\ \midrule
\textbf{BP2} & 1.04 & 0.99 & 1.00 & 0.99 & 1.00 \\ \midrule
\textbf{BP3} & 0.90 & 1.06 & 1.00 & 1.01 & 1.00 \\ \midrule
\textbf{BP4} & 0.97 & 1.05 & 1.00 & 0.98 & 1.00 \\ \midrule
\textbf{BP5} & 0.99 & 1.01 & 1.00 & 1.00 & 1.00 \\ \midrule
\textbf{BP6} & 0.99 & 1.00 & 1.00 & 1.01 & 1.00 \\ \midrule
\textbf{BP7} & 1.00 & 0.96 & 0.99 & 1.05 & 1.00 \\ \bottomrule \vspace{1cm}
\end{tabular}

\centering
\begin{tabular}{cccccc}
\toprule
\multicolumn{1}{l}{\textbf{}} &
  $\sigma\left( g g \to H_1 \right)$ &
  $\sigma\left( g g \to A_1 \right)$ &
  $\sigma\left( g g \to H_1  b \bar{b} \right)$ &
  $ B\left( {H_{1}} \to \tau \tau \right)$ &
  $ B\left( {A_{1}} \to \tau \tau \right)$ \\ \midrule
\textbf{BP1} & 15.99 & 56.61 & 4.41 & 3.32$ \times 10^{-5}$ & 1.81$ \times 10^{-4}$  \\ \midrule
\textbf{BP2} & 5.76 & 16.34  & 1.67  & 1.56$ \times 10^{-4}$ & 2.92$ \times 10^{-4}$  \\ \midrule
\textbf{BP3} & 8.66$ \times 10^{-4}$ & 1.84  & 2.40 & 8.86$ \times 10^{-6}$ & 2.50$ \times 10^{-1}$  \\ \midrule
\textbf{BP4} & 1.31$ \times 10^{-2}$ & 2.80$ \times 10^{-2}$ & 3.06$ \times 10^{-1}$ & 1.01$ \times 10^{-4}$ & 7.55$ \times 10^{-1}$  \\ \midrule
\textbf{BP5} & 1.31$ \times 10^{-4}$ & 4.70$ \times 10^{-1}$ & 7.29$ \times 10^{-2}$ & 1.29$ \times 10^{-3}$ & 3.11$ \times 10^{-1}$  \\ \midrule
\textbf{BP6} & 5.18 & 17.34 & 1.58 & 1.55$ \times 10^{-2}$ & 1.11$ \times 10^{-3}$  \\ \midrule
\textbf{BP7} & 2.43 & 9.92$ \times 10^{-1}$ & 1.14$ \times 10^{-1}$ & 2.28$ \times 10^{-2}$ & 3.34$ \times 10^{-5}$ \\ \bottomrule
\end{tabular}
\caption{A selection of seven benchmark points. All masses are given in GeV and cross sections in pb. These correspond to the lightest scalars found that respect all QFV, electroweak, Higgs and theoretical constraints (BP1 -- BP4), and to three early discovery/exclusion points that mostly approach the experimental bounds in \cref{fig:Atau} (BP5) as well as in \cref{fig:Htau}-right (BP6) and \cref{fig:Htau}-left (BP7).}
\label{tab:bench}
\end{table}

\begin{itemize}
	\item BP1 corresponds to the lightest CP-even BSM Higgs boson found in our scans with mass $m_{H_1} = 249~\mathrm{GeV}$. This point also corresponds the lightest charged Higgs scenario with $m_{H_1^\pm} = 101~\mathrm{GeV}$;
	\item BP2 represents the second-to-lightest CP-even and charged BSM Higgs bosons found in our anlysis with masses $m_{H_1} = 285~\mathrm{GeV}$ and $m_{H_1^\pm} = 146~\mathrm{GeV}$;
	\item BP3 and BP4 correspond to the lithest and next-to-lightest CP-odd Higgs boson found in our scan with masses $m_{A_1} = 161~\mathrm{GeV}$ and $206~\mathrm{GeV}$ respectively;
	\item BP5 offers an early discovery or early exclusion scenario in the $gg \to A_1 \to \tau \tau$ channel, $m_{A_1} = 338~\mathrm{GeV}$, where the signal strength was found to be the closest one to the CMS bound.
	\item BP6 corresponds to an early discovery/exclusion scenario in the $gg \to H_1  b \bar{b} \to \tau \tau  b \bar{b}$ channel, $m_{H_1} = 313~\mathrm{GeV}$, where the signal strength was found to be the closest one to the CMS bound.
	\item Last but not least, BP7 represents an early discovery/exclusion scenario in the $gg \to H_1 \to \tau \tau$ channel, $m_{H_1} = 353~\mathrm{GeV}$, where the signal strength was found to be the closest one to the CMS bound.
\end{itemize}
Note that the entire scalar spectrum in BP3, BP4 and BP5 is lighter than $1~\mathrm{TeV}$ and potentially at the reach of the LHC run-III. Furthermore, it is remarkable to note that the lightest charged Higgs in BP1 is allowed to be lighter than the SM Higgs boson while conforming with all experimental constraints. On the other hand, in BP1 and BP2 the heavy scalar masses $m_{H_2}$, $m_{A_2}$ and $m_{H_2^\pm}$ are larger than $4~\mathrm{TeV}$ while in BP6 and BP6 their masses are approximately $1.5~\mathrm{TeV}$ and $1.1~\mathrm{TeV}$. We also provide in \cref{tab:bench} both the production cross sections and the branching fractions calculated for each of the studied channels as well as the 3HDM-to-SM ratio of each of the five QFV observables in \cref{tab:err}. While the former are relevant for direct searches for new scalars at the LHC, the latter may be probed in flavour experiments.

An extended parameter space domain which relaxes the condition of exact Higgs alignment -- while maintaining the full agreement with the current LHC measurements regarding the properties of the lightest Higgs boson -- would no doubt show a larger excluded region by means of the measurements in the $gg\rightarrow H_1 b\bar{b} \rightarrow \tau\bar{\tau} b\bar{b}$ channel.

Other search channels might also be considered (such as searches in decays to top pairs) but we relegate those analyses for a future work where one would need to perform a more thorough scan of the model's parameter space. Our main goal here is to prove, with a simple example, that the model is of interest for the ongoing LHC searches.

\section{Conclusions}
\label{sec:conc}

We have presented a new BGL-like 3HDM model with symmetry-suppressed FCNC interactions mediated by neutral scalars, in a close similarity the well-known example of BGL 2HDM. We have applied the basic theoretical (unitarity, boundedness from below) and experimental (electroweak, Higgs and flavour) constraints and identified the domains of validity of the model and main phenomenological implications. Our analysis has enabled us to narrow down the allowed parameter space regions which simultaneously fit all theoretical and experimental constraints. We have also discussed the possibility of probing the model at the LHC via gluon fusion production of new CP-even and CP-odd Higgs bosons and subsequent decay into $\tau \bar \tau$ pairs as well as via an associated production of the new CP-even Higgs state and $b\bar b$ pair. In particular, we have observed that the BGL-like 3HDM offers a possibility for lighter than conventionally allowed non-standard scalars, at the reach of the LHC III. We have identified and described seven benchmark scenarios that can be used in experimental searches for Higgs partners at forthcoming LHC runs.

Our analysis determined the most sensitive flavour violation channels, and has revealed that the BGL-like mechanism induced by the $\mathrm{U(1)}\times\mathbb{Z}_2$ flavour symmetry is indeed responsible for the suppression of FCNCs rather than any accidental cancellation or any fine-tuning. Indeed, it results from the BGL nature of the 3HDM under consideration that the most stringent constraints on the model's parameter space are not the QFV ones but rather the Higgs physics observables from direct searches. As one of the possible future avenues, theoretical investigation of the proposed BGL-like mechanism can be continued towards generalisation of the lepton and neutrino sectors by implementing the $\mathrm{U(1)}\times\mathbb{Z}_2$ symmetry there and studying its consequences on LFV and neutrino mass generation.

\vspace*{10mm} {\bf Acknowledgements.}
This work is supported by the Center for Research and Development in Mathematics and Applications (CIDMA) through the Portuguese Foundation for Science and Technology (FCT - Funda\c c\~ao para a Ci\^encia e a Tecnologia), references UIDB/04106/2020 and, UIDP/04106/2020, and by national funds (OE), through FCT, I.P., in the scope of the framework contract foreseen in the numbers 4, 5 and 6 of the article 23, of the Decree-Law 57/2016, of August 29, changed by Law 57/2017, of July 19. This work is also supported by the Center for Theoretical and Computational Physics (CFTC) through the FCT contracts UIDB/00618/2020 and UIDP/00618/2020. We acknowledge support from the projects PTDC/FIS-PAR/31000/2017, CERN/FIS-PAR/0014/2019, CERN/FIS-PAR/0027/2019 and by HARMONIA project's contract UMO-2015/18/M/ST2/00518.
D.D.  thanks  the  Science  and  Engineering  Research  Board,  India  for  financial  support  through  grant  no.SRG/2020/000006.
R.P.~is supported in part by the Swedish Research Council grant, contract number 2016-05996, as well as by the European Research Council (ERC) under the European Union's Horizon 2020 research and innovation programme (grant agreement No 668679).
I.P.~acknowledges support from the Danmarks Frie Forskningsfonds (ProjectNo. 8049-00038B).

\appendix

\section{Generic off-alignment conditions}
\label{app:gen}

An alternative inversion procedure to the one used in our numerical calculations can be given. In particular, we provide in this appendix closed expressions for the quartic couplings $\lambda_{1,\ldots,10}$ and for the soft breaking mass terms $\mu_{12}^2$ and $\mu_{23}^2$ in terms of all physical scalar masses, mixing angles and VEVs.

Starting with the pseudoscalar sector, one can first invert (\cref{e.Ogamma2}) with respect to the elements of the ${B}^2_{P}$ matrix such that,
\begin{equation}
\label{eq:3HDM_Pseudo_relation}
\begin{gathered}
m_{\text{A}_1}^2 \cos ^2\left(\gamma _1\right)+m_{\text{A}_2}^2 \sin ^2\left(\gamma _1\right)) = \left( B^2_P \right)_{22}\,, \\
m_{\text{A}_2}^2 \sin \left(\gamma _1\right) \cos \left(\gamma _1\right) {\color{purple} -} m_{\text{A}_1}^2 \sin \left(\gamma _1\right) \cos \left(\gamma _1\right) = \left( B^2_P \right)_{32} \,, \\
m_{\text{A}_1}^2 \sin ^2\left(\gamma _1\right)+m_{\text{A}_2}^2 \cos ^2\left(\gamma _1\right) = \left( B^2_P \right)_{33}\,.
\end{gathered}
\end{equation}
Equating these to the corresponding quantities in \cref{eq:3HDMP_pseudo_intermedium}, one obtains three equations which can be resolved with respect to the potential parameters, for example, $\lambda_{10}$,
$\mu_{12}^2$ and $\mu_{23}^2$ in terms of two physical masses, $m_{A_1}$ and $m_{A_2}$, one mixing angle $\gamma_1$, the third soft mass parameter, $\mu_{13}^2$, and the Higgs VEVs (or, equivalently, mixing angles $\beta_{1,2}$ and $v$).
The results reads,
\begin{equation}
\begin{split}
\lambda_{10} = & \frac{\mu_{13}^2 v^2 v_{13}^2 +m_{\text{A}_1}^2 (v v_3 c_{\gamma_1} - v_1 v_2 s_{\gamma_1} )(v v_1 c_{\gamma_1} + v_2 v_3 s_{\gamma_1}) - m_{\text{A}_2}^2 (v_2 v_3 c_{\gamma_1} - v v_1 s_{\gamma_1})(v_1 v_2 c_{\gamma_1} + v v_3 s_{\gamma_1})}{2 v^2 v_1 v_3 v_{13}^2} \,,
\end{split}
\label{e.l10}
\end{equation}
\begin{equation}
\begin{split}
\mu_{23}^2 = & \frac{1}{v^2} \Bigg[ m_{\text{A}_2}^2 c_{\gamma_1} (v v_1 s_{\gamma_1} - v_2 v_3 c_{\gamma_1})
- m_{\text{A}_1}^2 s_{\gamma_1} \left(v_2 v_3 s_{\gamma_1} - v v_1 c_{\gamma_1}\right) \Bigg] \,,
\end{split}
\label{e.m23}
\end{equation}
\begin{equation}
\begin{split}
\mu_{12}^2 = & \frac{1}{v^2} \Bigg [ m_{\text{A}_1}^2 s_{\gamma_1} \left(v v_3 c_{\gamma_1} - v_1 v_2 s_{\gamma_1}\right)
- m_{\text{A}_2}^2 c_{\gamma_1} \left(v_1 v_2 c_{\gamma_1} + v v_3 s_{\gamma_1}\right) \Bigg] \,.
\end{split}
\label{e.m12}
\end{equation}

For the charged scalar masses one can use the expressions for $\lambda_{10}$, $\mu_{12}^2$ and $\mu_{23}^2$ found above and, solving the eigenvalue problem for \cref{e:BCrot}, extract the relations for other three parameters of the potential -- $\lambda_{7},\lambda_8$, and
$\lambda_9$ couplings -- in terms of the physical masses of the $H_{1,2}^{\pm}$ and $A_{1,2}$ states, the $\mu_{13}^2$ parameter, the
mixing angles $\gamma_{1,2}$ and the Higgs VEVs (or, equivalently, mixing angles $\beta_{1,2}$ and $v$). The result reads as
\begin{equation}
\begin{split}
\lambda_7 = & \dfrac{2}{v^2 v_1 v_2} \Bigg[ m_{\text{A}_2}^2 v_1 v_2 c_{\gamma_1}^2 + (m_{\text{A}_2}^2 - m_{\text{A}_1}^2) v v_3 c_{\gamma_1} s_{\gamma_1} +  m_{\text{A}_1}^2 v_1 v_2 s_{\gamma_1}^2 + m_{\text{C}_1}^2 s_{\gamma_2}(v v_3 c_{\gamma_2} - v_1 v_2 s_{\gamma_2})
 \\ & - m_{\text{C}_2}^2 c_{\gamma_2} (v_1 v_2 c_{\gamma_2} + v v_3 s_{\gamma_2}) \Bigg]
\\
\lambda_8 = & \dfrac{1}{v^2 v_1 v_3 v_{13}^2} \Bigg[\dfrac12 \Bigg( (m_{\text{A}_1}^2 + m_{\text{A}_2}^2) v_1 v_3 (v^2 - v_2^2) + (m_{\text{A}_1}^2 - m_{\text{A}_2}^2) \Big( v_1 v_3 (v^2 + v_2^2) c_{2 \gamma_1} + v v_2 (v_3^2 - v_1^2) s_{2 \gamma_1} \Big)  \Bigg)
\\ & - (m_{\text{C}_1}^2 - m_{\text{C}_2}^2) \Big(v_1 v_3 (v^2 + v_2^2) c_{2 \gamma_2} + v v_2 (v_1^2 - v_3^2) s_{2 \gamma_2}\Big) - (m_{\text{C}_1}^2 + m_{\text{C}_2}^2) v_1 v_3 (v^2 - v_2^2) - 2 v^2 v_{13}^2 \mu_{13}^2 \Bigg]
\\
\lambda_9 = & \frac{2}{v^2 v_2 v_3} \Bigg[ (m_{\text{A}_1}^2 - m_{\text{A}_2}^2) v v_1 c_{\gamma_1} s_{\gamma_1} + v_2 v_3 ( m_{\text{A}_1}^2 s_{\gamma_1}^2 -  m_{\text{A}_2}^2 c_{\gamma_1}^2) - m_{\text{C}_1}^2 s_{\gamma_2} (v v_1 c_{\gamma_2} + v_2 v_3 s_{\gamma_2})  \\ & + m_{\text{C}_2}^2 c_{\gamma_2} (v v_1 s_{\gamma_2} - v_2 v_3 c_{\gamma_2}) \Bigg]
\end{split}
\label{e:lam789}
\end{equation}

Finally, for the CP-even scalar masses, inverting \cref{e:msdiag} we get,
\begin{equation}\label{e:MS}
{M}_{S}^2  \equiv {\cal O}_\alpha^T  \cdot \begin{pmatrix}
m_h^2 & 0 & 0 \\
0& m_{H1}^2 & 0 \\
0 & 0 & m_{H2}^2 \\
\end{pmatrix} \cdot {\cal O}_\alpha\,,
\end{equation}
which enables us to solve for the remaining six quartic couplings
$\lambda_{1,2,3,4,5,6}$ in terms of the physical CP-even scalar masses, the Higgs VEVs and the mixing angles $\alpha_{1,2,3}$, as well as
the quartic couplings $\lambda_{7,8,9,10}$, and the soft parameters $\mu_{12}^2$ and $\mu_{23}^2$, obtained above for the CP-odd and
charged scalar sectors. The final expressions can be expressed as
\begin{equation}
\begin{split}
\lambda_1 =& \dfrac{1}{2 v^2 v_1^3} \Bigg[v^2 v_3 \mu_{13}^2 + m_h^2 v^2 v_1 c_{\alpha_1}^2 c_{\alpha_2}^2 + v_2 \Big( (m_{\text{A}_1}^2 - m_{\text{A}_2}^2) v v_3 c_{\gamma_1} s_{\gamma_1} - v_1 v_2 (m_{\text{A}_1}^2 s_{\gamma_1}^2 + m_{\text{A}_2}^2 c_{\gamma_1}^2) \Big) \\ & v^2 v_1 \Big( m_{\text{H}_1}^2 (s_{\alpha_2} s_{\alpha_3} + s_{\alpha_1} c_{\alpha_2} c_{\alpha_3} )^2 - m_{\text{H}_2}^2 (s_{\alpha_2} c_{\alpha_3} - s_{\alpha_1} c_{\alpha_2} s_{\alpha_3})^2 \Big)\Bigg]
\\
\lambda_2 =& \dfrac{1}{2 v^2 v_2^2} \Bigg[m_h^2 v^2 s_{\alpha_1}^2 - v_{13}^2 (m_{\text{A}_1}^2 s_{\gamma_1}^2 + m_{\text{A}_2}^2 c_{\gamma_1}^2) + v^2 c_{\alpha_1}^2 (m_{\text{H}_1}^2 c_{\alpha_3}^2 + m_{\text{H}_2}^2 s_{\alpha_3}^2)\Bigg]
\\
\lambda_3 =& \dfrac{1}{2 v^2 v_3^3} \Bigg[ v^2 \Bigg( m_h^2 v_3 c_{\alpha_1}^2 s_{\alpha_2}^2 + v_1 \mu_{13}^2 + v_3 \Big(m_{\text{H}_1}^2 (s_{\alpha_1} s_{\alpha_2} c_{\alpha_3} - c_{\alpha_2} s_{\alpha_3})^2
+ m_{\text{H}_2}^2 (s_{\alpha_1} s_{\alpha_2} s_{\alpha_3} + c_{\alpha_2} c_{\alpha_3})^2 \Big) \Bigg) \\ &
- v_2 \Big( (m_{\text{A}_1}^2 - m_{\text{A}_2}^2) v v_1 c_{\gamma_1} s_{\gamma_1} + (m_{\text{A}_1}^2 s_{\gamma_1}^2 + m_{\text{A}_2}^2 c_{\gamma_1}^2)v_2 v_3  \Big) \Bigg]
\end{split}
\label{e:lam1to3}
\end{equation}
%%%%%
\begin{equation}
\begin{split}
\lambda_4 = & \dfrac{1}{v^2 v_1 v_2} \Bigg[ m_h^2 v^2 c_{\alpha_1} s_{\alpha_1} c_{\alpha_2} + (m_{\text{A}_1}^2 - m_{\text{A}_2}^2) v v_3 c_{\gamma_1} s_{\gamma_1} - (m_{\text{A}_1}^2 s_{\gamma_1}^2 + m_{\text{A}_2}^2 c_{\gamma_1}^2) v_1 v_2 \\& + 2 m_{\text{C}_1}^2 s_{\gamma_2} (v_1 v_2 s_{\gamma_2} - v v_3 c_{\gamma_2}) + 2 m_{\text{C}_2}^2 c_{\gamma_2} (v_1 v_2 c_{\gamma_2} + v v_3 s_{\gamma_2}) \\&
-v^2 c_{\alpha_1} \Big(m_{\text{H}_1}^2 c_{\alpha_3} (s_{\alpha_1} c_{\alpha_2} c_{\alpha_3} + s_{\alpha_2} s_{\alpha_3})
+m_{\text{H}_2}^2 s_{\alpha_3} (s_{\alpha_1} c_{\alpha_2} s_{\alpha_3} - s_{\alpha_2} c_{\alpha_3})
\Big)	
\Bigg]
\\
\lambda_5 =& \dfrac{1}{v^2 v_1 v_3 v_{13}^2} \Bigg[m_{\text{C}_1}^2 \Big(v_1 v_3 (v^2 - v_2^2) + v_1 v_3 (v^2 + v_2^2) c_{2 \gamma_2} + v v_2 (v_3^2 - v_1^2) s_{2 \gamma_2}\Big) \\&
m_{\text{C}_2}^2 \Big(v_1 v_3 (v^2 - v_2^2) - v_1 v_3 (v^2 + v_2^2) c_{2 \gamma_2} + v v_2 (v_1^2 - v_3^2) s_{2 \gamma_2}\Big) \\&
v^2 v_{13}^2 \Bigg(\mu_{13}^2 + 4 m_h^2 c_{\alpha_1}^2 s_{2 \alpha_2} - (m_{\text{H}_1}^2 - m_{\text{H}_2}^2)(c_{2 \alpha_1} - 3) c_{2 \alpha_3} s_{2 \alpha_2} + 4 (m_{\text{H}_2}^2 - m_{\text{H}_1}^2) s_{\alpha_1} c_{2 \alpha_2} s_{2 \alpha_3} \\& -2 (m_{\text{H}_1}^2 + m_{\text{H}_2}^2) c_{\alpha_1}^2 s_{2 \alpha_2} \Bigg) \Bigg]
\end{split}
\label{e:lam4to5}
\end{equation}
%%%%%
\begin{equation}
\begin{split}
\lambda_6 =& \dfrac{1}{v^2 v_2 v_3} \Bigg[ v_1 v s_{\gamma_1} c_{\gamma_1} (m_{\text{A2}}^2-m_{\text{A1}}^2)
-v_2 v_3 ( m_{\text{A1}}^2 s^2_{\gamma_1}
+ m_{\text{A2}}^2 c^2_{\gamma_1} )
+2 s_{\gamma_2} m_{\text{C1}}^2 (v_2 v_3 s_{\gamma_2}+v v_1 c_{\gamma_2}) \\&
+2 c_{\gamma_2} m_{\text{C2}}^2 (v_2 v_3 c_{\gamma_2}-v v_1 s_{\gamma_2})
+v^2 c_{\alpha_1} \Big(c_{\alpha_3} m_{\text{H1}}^2 (c_{\alpha_2} s_{\alpha_3}-s_{\alpha_1} s_{\alpha_2} c_{\alpha_3}) \\& -s_{\alpha_3} m_{\text{H2}}^2 (s_{\alpha_1} s_{\alpha_2} s_{\alpha_3}+c_{\alpha_2} c_{\alpha_3})\Big)
+m_h^2 v^2 c_{\alpha_1} s_{\alpha_1} s_{\alpha_2} \Bigg]
\end{split}
\label{e:lam6to6}
\end{equation}

In essence, a generic, off-alignment scan, can be performed by exchanging $\mu_{12}^2$, $\mu_{23}^2$ and $\lambda_{10}$ for $m_{A1}$, $m_{A2}$ and the mixing angle $\gamma_1$ using \cref{e.l10,e.m23,e.m12}. The remaining nine quartic couplings, as discussed above, can be expressed in terms of five physical masses (three CP-even scalars, and two charged scalars), the remaining soft-breaking parameter, $\mu_{13}^2$, and four mixing angles (three in the CP-even sector and one in the charged scalar sector).

%%%%%%%%%%%%%%%%%%%%%%%%%%%%%%%%%%%%%%%%%%%%%%%%%%%%%%%%%%%%%
\bibliographystyle{JHEP}
\bibliography{Ref.bib}
\end{document}